\documentclass[aip,cha,preprint,numerical]{revtex4-1}
\usepackage{mathtools}
\usepackage{amsmath}
\usepackage{amssymb}
\usepackage{graphicx}
\usepackage{dcolumn}
\usepackage{bm}
\usepackage{natbib}		
\usepackage{color}

\newcommand {\e} {\varepsilon}
\newcommand {\eps} {\varepsilon}

\newcommand{\avr}[1]{\left\langle #1 \right\rangle}

\begin{document}
\title{Attractor-repeller collision and the heterodimensional dynamics}

\author{Vladimir Chigarev}
\email{vchigarev@hse.ru}
\affiliation{National Research University Higher School of Economics, \\
             25/12 Bolshaya Pecherskaya Ulitsa, 603155 Nizhny Novgorod, Russia}
\author{Alexey Kazakov}
\email{kazakovdz@yandex.ru}
\affiliation{National Research University Higher School of Economics, \\
             25/12 Bolshaya Pecherskaya Ulitsa, 603155 Nizhny Novgorod, Russia}
\author{Arkady Pikovsky}
\email{pikovsky@uni-potsdam.de}
\affiliation{Institute of Physics and Astronomy, University of Potsdam,
Karl-Liebknecht-Str. 24/25, 14476 Potsdam-Golm, Germany}

\date{\today}

\begin{abstract}
We study the heterodimensional dynamics in a simple map on a three-dimensional torus. This map consists of a two-dimensional driving Anosov map and a one-dimensional driven M\"obius map, and demonstrates the  collision of a chaotic attractor with a chaotic repeller if parameters are varied. We explore this collision by following tangent bifurcations of the periodic orbits, and establish a regime where periodic orbits with different numbers of unstable directions coexist in a chaotic set. For this situation, we construct a heterodimensional cycle connecting these periodic orbits. Furthermore, we discuss properties of the rotation number and of the nontrivial Lyapunov exponent at the collision and in the heterodimensional regime.
\end{abstract}

\maketitle

\begin{quotation}
Hyperbolic dynamics is the most perfect form of chaos, with many nice mathematical and physical properties. However, in realistic systems one often deals with not so perfect situations, where hyperbolicity is broken. One particular case of broken hyperbolicity is that of heterodimensional dynamics, where inside a chaotic set there are orbits with different numbers of unstable directions. In this paper we present a particular mechanism of appearance of heterodimensional dynamics via a collision of a chaotic attractor and a chaotic repeller. We follow this transition by determining bifurcations of the periodic orbits embedded in the attractor and the repeller. We also demonstrate that heterodimensional dynamics has implications for the distribution of finite-time Lyapunov exponents.
\end{quotation}

\section{Introduction}

Hyperbolic chaotic sets are ideal objects with nice dynamical and statistical properties, however, they are quite rare in real applications. Hyperbolicity can be broken in many ways, and one particular case, that has attracted much attention recently, is that of \textit{heterodimensional dynamics}, where different orbits inside a chaotic set have different numbers of positive Lyapunov exponents. In particular, periodic orbits embedded in chaos can have different numbers of stable and unstable directions. Remarkably, conclusions about the robustness of such a situation can be drawn just from the existence of so-called heterodimensional cycles
\cite{BonattiDiaz96} -- trajectories, connecting periodic orbits with different dimensions of stable and unstable manifolds.

One of the first examples of hyperbolicity breaking associated with the existence of heterodimensional cycles was presented by Abraham and Smale \cite{AbrahamSmale70}. The persistence of non-transversal intersections of invariant manifolds in such cycles was discovered and studied by D\'iaz and his collaborators \cite{DiazRocha92, diaz1995robust, diaz1995persistence}. The mathematical theory for maps with heterodimensional cycles of co-index one (where the difference between the dimensions of the unstable manifolds of the pair of saddle orbits connected by these cycles is one) was developed by Bonatti and D\'iaz \cite{BonattiDiaz96, BonattiDiaz08}, where the authors proved the $C^1$-robustness of heterodimensional cycles. A more general higher smoothness version of this result has been recently obtained by Li and Turaev~\cite{LiTuraev2021}. The results of Ref.~\onlinecite{LiTuraev2021} imply the existence of the heterodimensional dynamics caused by heterodimensional cycles in an appropriate two-parameter family that unfolds an initial heterodimensional cycle.

In this paper we explore a mechanism for the appearance of heterodimensional dynamics through a collision of a simple chaotic attractor with a simple chaotic repeller (both these objects are hyperbolic prior to the collision). Such a collision occurs naturally at the breakout of chaotic phase synchronization~\cite{Pikovsky-Osipov-Rosenblum-Zaks-Kurths-97} (although in realistic situations one can hardly assume hyperbolicity). This collision is an extension of a simple saddle-node (tangent) bifurcation, responsible for the loss of synchrony of periodic driven oscillators, to the case of chaotic amplitude dynamics. We formulate the basic system as a chaotic hyperbolic subsystem (Anosov torus map), driving a circle map (Section \ref{sec:bm}).

In the context of the chaotic phase synchronization applications~\cite{Pikovsky-Osipov-Rosenblum-Zaks-Kurths-97}, the chaotic subsystem corresponds to the amplitude dynamics, and the circle map describes the driven phase dynamics. A stable regime of chaotic phase synchronization corresponds to the existence of a chaotic attractor on which the phase variations are bounded, roughly ``in phase'' with the driving. Since in the considered model the amplitude dynamics is invertible, there also exists a chaotic repeller (an attractor in the inverted time), roughly ``in antiphase'' with the driving. The dimensions of the stable and unstable manifolds of orbits on the attractor and on the repeller follow from one unstable and one stable directions of the Anosov map, and from the stability (instability) for the attractor (repeller) in the phase direction. As a parameter responsible for the ``frequency mismatch'' varies, the attractor and the repeller collide (Section~\ref{sec:parc}).

We show that this collision can be best traced through tangent bifurcations of periodic orbits embedded in chaos (Section \ref{sec:cupo}). We argue that beyond the collision the attractor and the repeller overlap, and one chaotic set appears containing orbits with different dimensions of unstable manifolds. For two-dimensional reversible maps such phenomenon was studied in Refs.~\onlinecite{gonchenko2017phenomenon, Kazakov2019, kazakov2020merger} where it was also shown that after this a new type of chaotic dynamics -- the so-called \textit{mixed dynamics} appears. The mathematical theory of mixed dynamics was developed in Refs.~\onlinecite{gonchenko2016reversible, Gonchenko2017, turaev2021criterion}; recent examples are given in Refs. \onlinecite{emelianova2019intersection, ariel2020conservative, gonchenko2020three, kuznetsov2020chaplygin, emelianova2020third, emelianova2021emergence}.

Furthermore, we show in Section \ref{sec:hc} that right beyond the attractor--repeller collision a heterodimensional cycle appears connecting a pair of saddle orbits from the former attractor and repeller. This implies, according to the Li and Turaev theorem \cite{LiTuraev2021}, that arbitrarily close to this cycle other heterodimensional cycles exist. Thus, one can say that the heterodimensional dynamics caused by the observed heterodimensional cycle is robust, i.e., it persists under variations of the parameter values.

\section{Model}
\label{sec:bm}

The basic model we study in this paper is a chaotically-driven circle map. Because we want to have invertable equations, the driving chaotic map must be at least two-dimensional. We take a standard Anosov torus map (a.k.a. Arnold cat map) defined on a 2D-torus, which has nice properties: it is hyperbolic, and the stable and unstable manifolds can be easily found as straight lines on the torus. As a circle map we take a M\"obius map~\cite{Marvel-Mirollo-Strogatz-09, Gong-Toenjes-Pikovsky-20, Chigarev-Kazakov-Pikovsky-20, Chigarev-Kazakov-Pikovsky-21}, which is readily invertible (for convenience, we summarize relevant properties of M\"obius map
in appendix \ref{sec:app}). Thus, the system under investigation is an invertible map on 3D-torus $0\leq t,s,x <1$:
\begin{subequations}
\begin{align}
t_{n+1}&=2t_n+s_n\pmod{1}\;,\label{eq:bma1}\\
s_{n+1}&=t_n+s_n\pmod{1}\;,\label{eq:bma2}\\
x_{n+1}&=x_n+c+\mu\sin(2\pi t_n+\alpha)-\frac{1}{\pi}\arctan
\left(\frac{\e \sin 2\pi x_n}{1+\e\cos 2\pi x_n}\right)\pmod{1}.
\label{eq:bmm}
\end{align}
\label{eq:bm}
\end{subequations}
Here Eqs.~\eqref{eq:bma1}--\eqref{eq:bma2} describe the driving Anosov map; it has no free parameters. Equation \eqref{eq:bmm} is the driven M\"obius map. The free M\"obius map (i.e. with $\mu=0$) has two parameters $\e$ and $c$. Parameter $\e$ determines how close is this map to a circle shift, which is realized for $\e=0$. In the limit $\e\to 1$ almost the whole circle is mapped to a small neighborhood of one point on it. Parameter $c$ is an additive one, it defines in a natural way the rotation number of the free M\"obius map.

In some situations below, e.g. at calculations of the rotation number,
it is convenient to lift map \eqref{eq:bmm} from the unit circle to
the real line, in this case one just drops the operation  $\pmod{1}$. We will denote the corresponding variable on the real line as $\tilde{x}_n$.

In the context of chaotic phase synchronization~\cite{Pikovsky-Osipov-Rosenblum-Zaks-Kurths-97}, where a periodically driven chaotic attractor with a well-defined phase variable is studied, system \eqref{eq:bm} and the parameters have the following interpretations. The Anosov map \eqref{eq:bma1}--\eqref{eq:bma2} describes chaos of the amplitude variables of  the attractor, while variable $x$ in Eq.~\eqref{eq:bmm} corresponds to the phase. Parameter $\mu$ describes ``internal coupling'' between the amplitude and the phase; it governs phase diffusion and is related to the level of coherence of free chaotic oscillations (larger values of $\mu$ correspond to a stronger phase diffusion, small values of $\mu$ mean almost regular phase rotations). Particularities of this internal coupling depend on the additional phase shift $\alpha$. Terms $c$ and $\e$ describe the effect of the external periodic force on the chaotic attractor, their meaning is the same as in the context of a circle map reduction for forced periodic oscillations: $c$ is roughly proportional to the frequency mismatch, while $\e$ describes the strength of the forcing.

Below we will describe attractors and repellers and their metamorphoses in dependence on the parameters. Parameter $\e$ will be mainly fixed to $\e=0.4$, due to the following reasons.  Nontrivial dynamics of $x$ disappears in the limit $\e\to 0$, where  map \eqref{eq:bmm}
is just the forced circle shift. On the other hand, in the limit $\e\to 1$  map \eqref{eq:bmm} becomes singular. While one does not expect
any qualitative dependence of the regimes described below in $\e$ (see Fig.~\ref{fig:arbd}), mostly convenient for the
numerical analysis and especially for visualization are ``moderate'' values of parameter $\e$.

\section{Phenomenology of the attractor-repeller collision}
\label{sec:parc}

\begin{figure}
\includegraphics[width=\columnwidth]{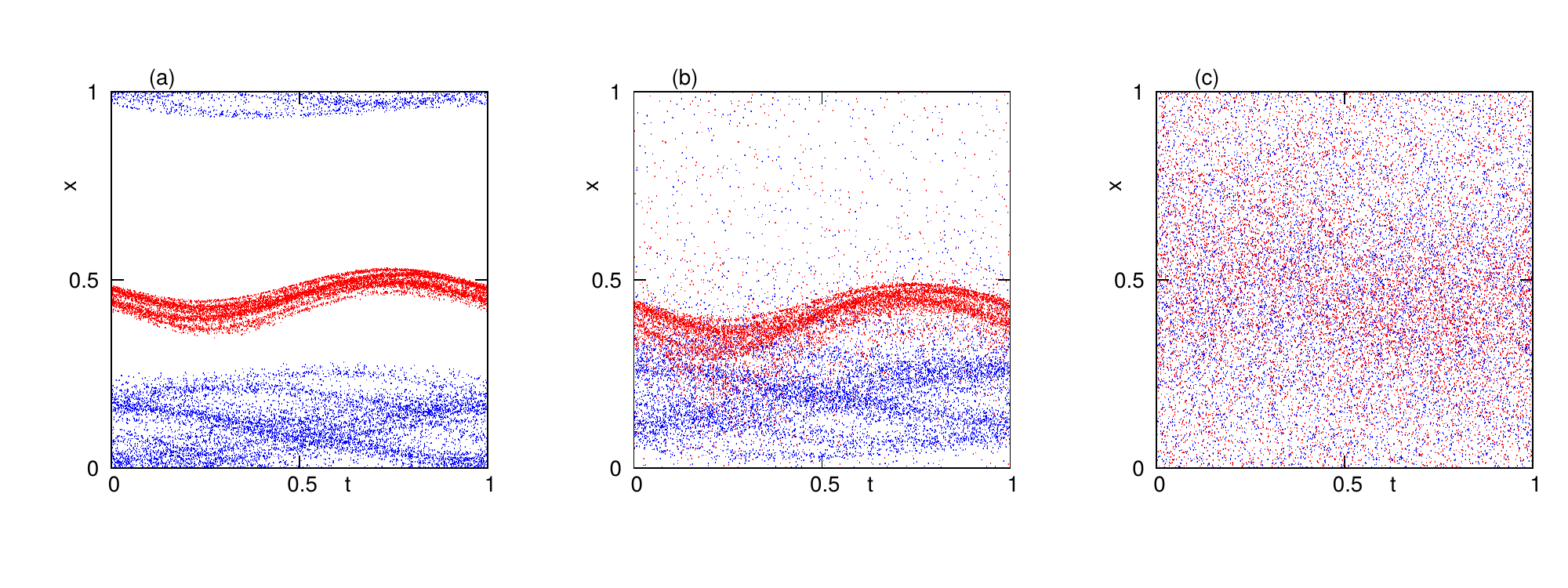}
\caption{Projections of the attractor (blue dots) and the repeller (red dots) for $\e=0.4$, $\alpha=0$ and $\mu=0.1$. Panel (a): $c=0.05$, the attractor and the repeller are separated;
panel (b): $c=0.1$, the attractor and the repeller overlap, but their measures are quite different; panel (c): $c=0.4$, the measures of the overlapping attractor and repeller are quite similar.}
\label{fig:ar}
\end{figure}

\subsection{Attractor and repeller}
We illustrate the chaotic attractor and the chaotic repeller in system \eqref{eq:bm} in Fig.~\ref{fig:ar}. We depict long trajectories forward and backward in time,
after discarding initial transients.
These trajectories give impression on the invariant measures of the attractor and of the repeller.  In the situation depicted in panel (a) these sets are separated, so that the attractor has its basin, and, in the forward time, the repeller serves as a chaotic saddle, close to which the long-lived
transient dynamics occurs~\cite{Lai-Tel-11}. For the backward in time dynamics, these sets
exchange their roles. Panels (b) and (c) show the case of overlapped attractor and repeller, where
both sets are dense in the whole phase space (3D torus). However, the invariant measures of the attractor and
the repeller  are quite different in the case of panel (b), while for the case of panel (c)
these measures nearly coincide. A transition from the separated attractor and repeller
(Fig.~\ref{fig:ar}a) to the overlapped chaotic set (Figs.~\ref{fig:ar}b,c) is
called \textit{attractor-repeller collision}~\cite{Pikovsky-Osipov-Rosenblum-Zaks-Kurths-97}.

\subsection{Rotation number and the Lyapunov exponent}

It is convenient to characterize the regimes in system \eqref{eq:bm} by means of the rotation
number and the Lyapunov exponent, both calculated for the driven variable $x$.

The rotation number, both for the free and for the driven M\"obius map, is defined as the mean velocity of rotations around the circle, using the lifted map:
\begin{equation}
\rho=\lim_{n\to\infty}\frac{\tilde{x}_{n}-\tilde{x}_0}{n}\;.
\label{eq:rndef}
\end{equation}
It is clear that only the interval of values $-\frac{1}{2}\leq c<\frac{1}{2}$ is of interest, because $\rho(c\pm 1)=\rho(c)\pm 1$.

Another quantity we will calculate in numerical simulations with system \eqref{eq:bm}, is the Lyapunov exponent (LE) in the $x$-direction.
Indeed, because \eqref{eq:bm} is a skew system with driving variables $(s,t)$, the three Lyapunov exponents are
combined from the two Lyapunov exponents of the Anosov cat map \eqref{eq:bma1}, \eqref{eq:bma2} and of the Lyapunov exponent
in the $x$ direction from \eqref{eq:bmm}. A direct calculation of the derivative of the r.h.s. of \eqref{eq:bmm}
\[
\frac{d x_{n+1}}{d x_n}=\frac{1-\e^2}{1+2\e\cos 2\pi x_n+\e^2}
\]
yields the expression
 $\lambda=\avr{\log \frac{d x_{n+1}}{dx_n}}=\avr{\log \frac{1-\e^2}{1+2\e\cos 2\pi x+\e^2}}$. Here the average should be taken according
 to the invariant measure on the attractor; due to ergodicity it can be practically calculated as the time average.

\begin{figure}
\includegraphics[width=0.6\columnwidth]{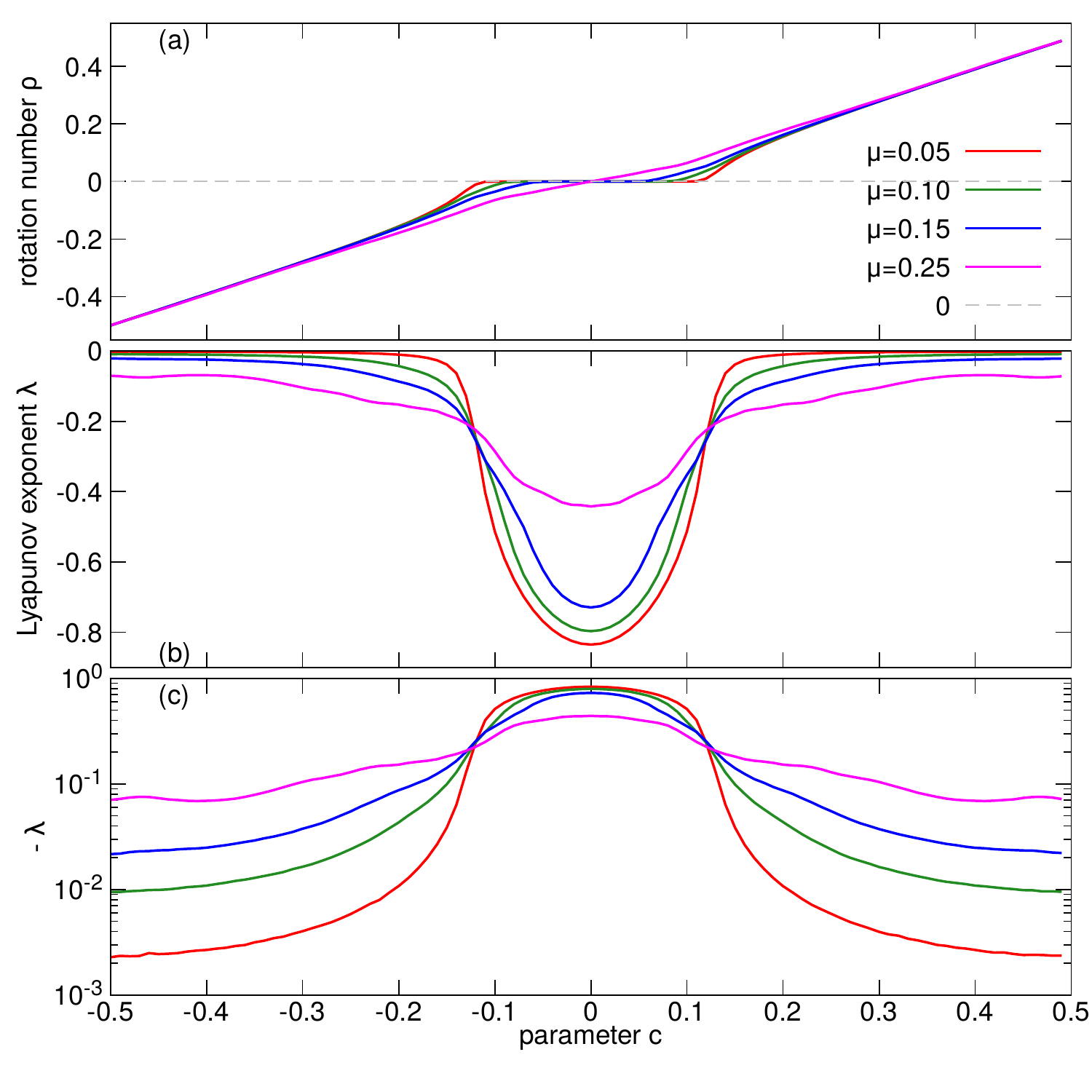}
\caption{Dependence of the rotation number (panel (a)) and the Lyapunov exponent (panel (b)) on parameter $c$ for $\e=0.4$ and $\alpha=0$, for several values of $\mu$. To reveal that the LE $\lambda$ is negative (although rather close to zero in some domain) in the whole range of parameters, we show $-\lambda$ in the logarithmic
scale in panel (c).}
\label{fig:rl}
\end{figure}

\begin{figure}
\includegraphics[width=0.6\columnwidth]{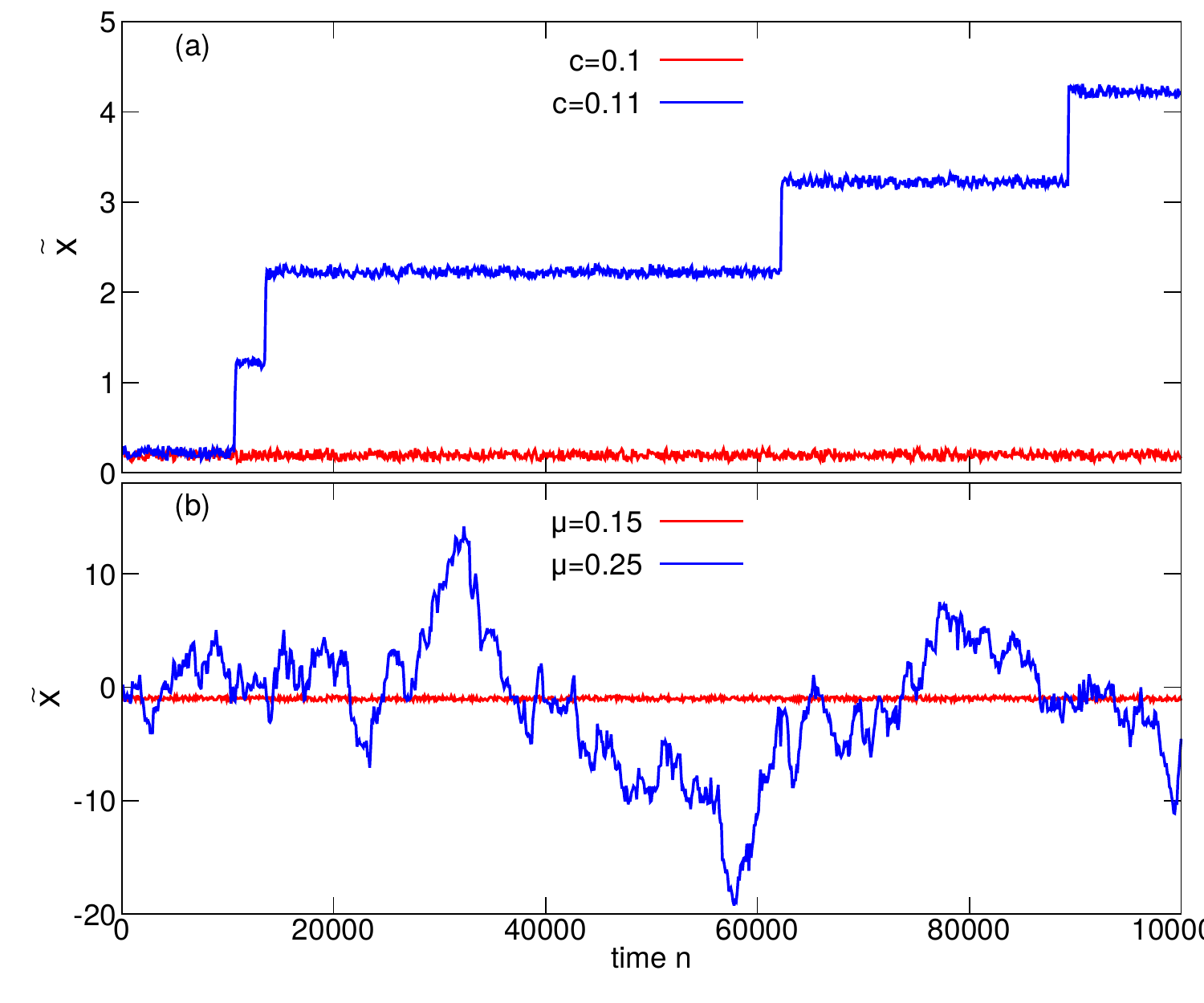}
\caption{Time series $\tilde{x}_n$ of the lifted map \eqref{eq:bmm} with $\e=0.4$ and $\alpha=0$. Panel (a): fixed forcing $\mu=0.04$ and two values of
parameter $c$, one with zero rotation number, and one where variable $\tilde x$ continuously grows.
Panel (b): symmetric case $c=0$. Here for $\mu=0.15$ there are separated attractor and repeller, and variable $\tilde x$ remains bounded.
For $\mu=0.25$, there attractor and repeller have already
collided and the evolution of $\tilde x$ represents unbounded diffusion.}
\label{fig:arcx}
\end{figure}

We illustrate application of the rotation number and of the nontrivial Lyapunov exponent to the characterization of the attractor-repeller collision in Fig. \ref{fig:rl}. It shows profiles of the quantities $\rho$ and $\lambda$ in dependence on parameter $c$ for fixed other parameters of the system. Both quantities are calculated in forward time (i.e. they characterize the attractor). The domain where $\rho=0$ is that of separated attractor and repeller (like in Fig.~\ref{fig:ar}a), because here variations of $\tilde{x}_n$ are bounded. States with $\rho\neq 0$ correspond to the overlapped attractor and repeller (like in Fig.~\ref{fig:ar}b,c). A separate consideration should be performed at the symmetric case $c=0$, where the rotation number vanishes due to statistical symmetry $x\leftrightarrow -x$, although the attractor and the repeller can overlap. We illustrate this in Fig.~\ref{fig:arcx}, where we show in panel (a) trajectories of $\tilde{x}_n$ in the asymmetric case $c\neq 0$, and in panel (b) in the symmetric case $c=0$.
In the latter situation, an overlap of the attractor and repeller can be detected by following the difference between maximal and minimal values of $\tilde x_n$ during a long time interval. As panel (b) of  Fig.~\ref{fig:arcx} illustrates, this difference takes large values (definitely larger than one) for overlapping attractor and repeller, and remains small if they do not overlap.

\begin{figure}
\includegraphics[width=0.4\columnwidth]{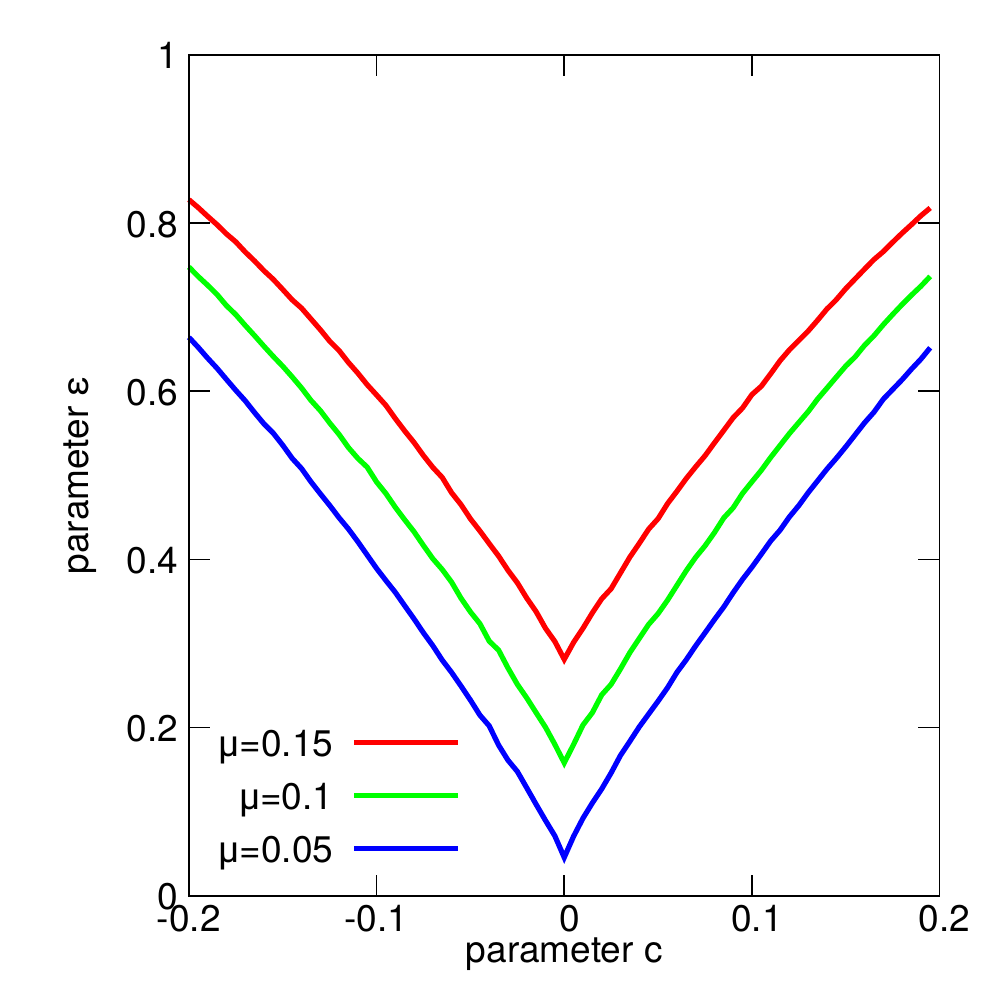}
\caption{Domains of existence of attractor and repeller on the $(c,\e)$-parameter plane (what corresponds to a usual representation of Arnold tongues), for several values of parameter $\mu$ and $\alpha=0$. The attractor and the repeller are separated above the curves, where the rotation number $\rho$ vanishes.}
\label{fig:arbd}
\end{figure}

Calculations of the Lyapunov exponent are presented in Fig.~\ref{fig:rl}b. This quantity remains negative in all cases, the only difference is in the level of the LE: for nearly symmetric overlapping attractor and repeller it is close to zero, while in the non-overlapping case its value is of order one. This property is different from the corresponding feature of an autonomous circle map: there inside Arnold tongues the LE is negative, while it vanishes at the quasiperiodic dynamics outside of the Arnold tongues. For the autonomous M\"obius map, which either possesses only one Arnold tongue  or otherwise is conjugated to a circle shift (see discussion in appendix \ref{sec:app}), the LE vanishes exactly in the whole interval of parameters outside the Arnold tongue. We will further discuss this property of the chaotically driven M\"obius map in Section \ref{sec:cupo} below.

Adopting the criterion for the overlap of the attractor and repeller as described above, we calculate in Fig.~\ref{fig:arbd} an analog of the Arnold tongue on plane of parameters $(c,\eps)$ for the system \eqref{eq:bm}. As discussed above, in the context of the phase synchronization theory, these parameters correspond to ``frequency mismatch'' (parameter $c$) and the ``amplitude of the force'' (parameter $\varepsilon$) for a periodically driven chaotic system, internal phase irregularity of which is described by parameter $\mu$. One can see that for large values of parameter $\mu$, larger values of the forcing are needed to ``phase synchronize'' the system, i.e., to ensure that the attractor and the repeller do not overlap and the rotation number vanishes.

\section{Collision of attractor and repeller in terms of periodic orbits}
\label{sec:cupo}

\subsection{Bifurcations of periodic orbits}
\label{sec:bpo}

In descriptions of transitions of chaotic sets, a common approach is to
follow the corresponding transitions of saddle periodic orbits, embedded in chaos. In the case of attractor-repeller collision in the skew system \eqref{eq:bm}, this method corresponds to considering bifurcations of the M\"obius map \eqref{eq:bmm}, driven by periodic orbits of the Anosov subsystem~\eqref{eq:bma1}--\eqref{eq:bma2}.

These bifurcations are rather simple, because the M\"obius map has a special property distinguishing it from generic circle maps: any iteration of a M\"obius map is again a M\"obius map (although with different parameters, see appendix \ref{sec:app}). Thus, for any period-$m$ orbit in the driving $(t_n,s_n)$-subsystem, the map  $x_n\to x_{n+m}$  according to \eqref{eq:bm} is a M\"obius map. Because the latter map is autonomous, it either
\begin{itemize}
\item{(i)} possesses a pair of stable and unstable fixed points (inside the basic ``synchronization region'', where the rotation number is an integer);
\item{or (ii)} can be transformed by a smooth transformation $x\to y$ to the circle shift $y_{n+m}=y_n+m\rho$, where $\rho$ is the rotation number according to \eqref{eq:rndef} (see appendix \ref{sec:app} for details). This number smoothly depends on the parameters, so that there are no windows of periodicity (no higher-order Arnold tongues).
\end{itemize}

\begin{figure}
\includegraphics[width=0.65\columnwidth]{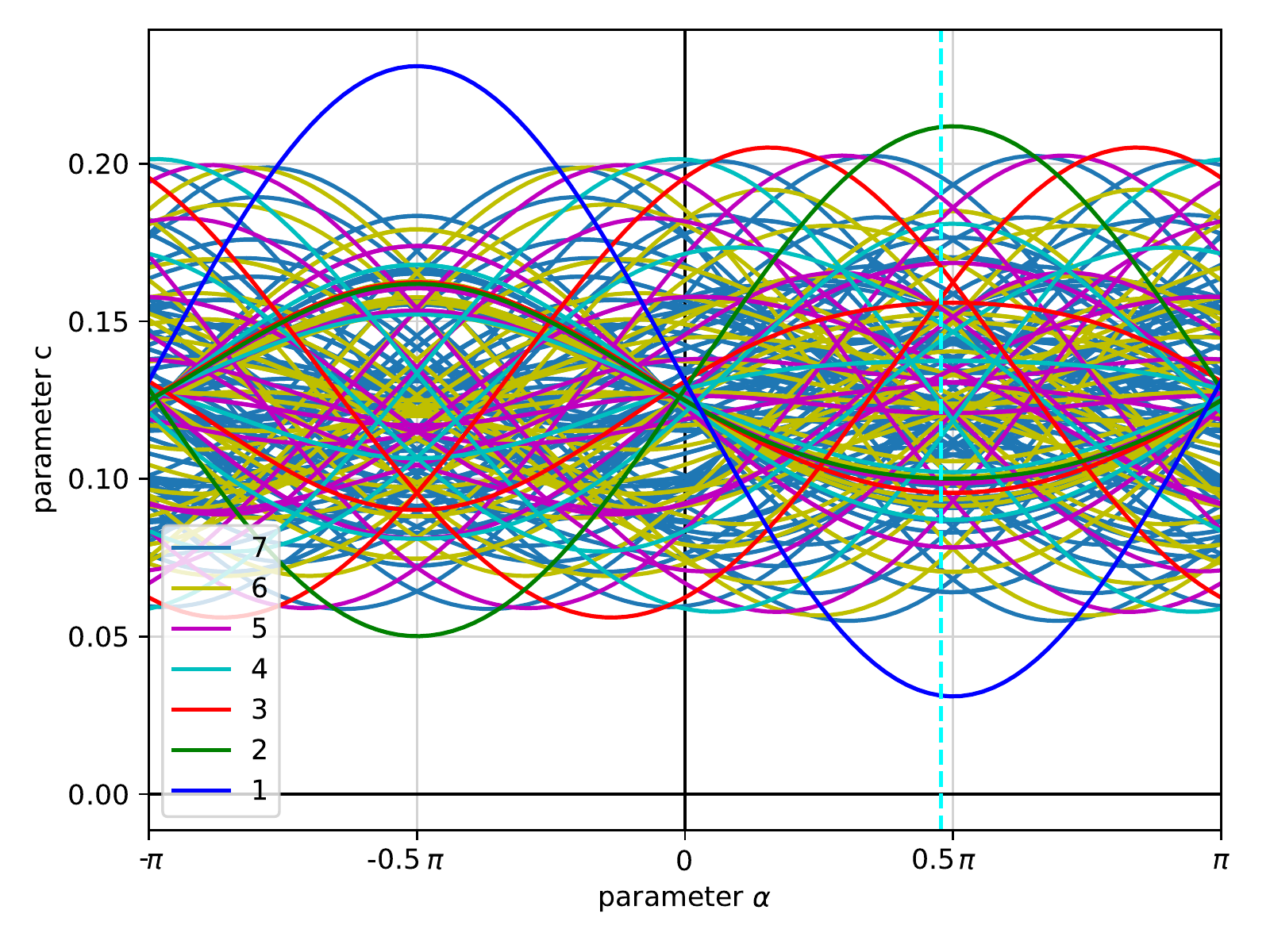}
\caption{Curves for saddle-node bifurcations in map \eqref{eq:bm} for all periodic orbits of the Anosov map with periods up to $7$ (curves for the periodic orbits
of the same period have the same color, as indexed on the panel),
on the plane of parameters $(c,\e)$ for $\e=0.4$ and $\mu=0.1$. The ``envelopes''
of these curves define border between regimes 1 and 3 (small values of $c$)
and between regimes 2 and 3 (large values of $c$).}
\label{fig:pbd}
\end{figure}

This feature of the M\"obius map means that for any driving periodic orbit $(t_n,s_n)$, in Eq.~\ref{eq:bmm} there is just one possible saddle-node (tangent) bifurcation which separates regimes (i) and (ii) above.
In Fig.~\ref{fig:pbd} we show these bifurcation curves in the plane of parameters $(c,\alpha)$ for fixed $\e=0.4$ and $\mu=0.1$.

One can see from Fig.~\ref{fig:pbd} that for every value of parameter $\alpha$ there is a range of critical positive values of parameter $c$, $0<c_{1}(\alpha,\mu,\e)\leq c \leq c_{2}(\alpha,\mu,\e)$, at which different periodic orbits bifurcate (the corresponding interval of negative values of $c$ is $c_3(\alpha,\mu,\e)\leq c\leq c_4(\alpha,\mu,\e)<0$). We will call these ranges of $c$ \textit{transitional regions}. Thus, the system demonstrates three dynamical states:
\begin{enumerate}
\item \textbf{Small $|c|$:} Separated attractor and repeller exist for $c_4(\alpha,\mu,\e) < c < c_1(\alpha,\mu,\e)$. On these sets the $x$-coordinates on the attractor and repeller are functions of $(t,s)$. These functions are expected to be relatively smooth for large $\e$ and are non-smooth (fractals) for small $\e$. All other points of the phase space belong to basins of the attractor (or of the repeller, if one iterates backward in time). The rotation number $\rho$ is zero.
\item \textbf{Large $|c|$:} Attractor and repeller overlap and possess no isolated periodic orbits for $c>c_2(\alpha,\e,\mu)$ and $c<c_3(\alpha,\e,\mu)$. Thus, there are no hyperbolic sets for this range of parameters. Evolution of variable $x$ for each periodic trajectory of the Anosov map is described by a superposition of M\"obius maps, which results in a M\"obius map that is smoothly conjugate to a circle shift. This means that the full system \eqref{eq:bm} possesses no isolated periodic orbits. The rotation number $\rho$ is non-zero. It is instructive to discuss the Lyapunov exponent in this region. According to calculations shown in Figs.~\ref{fig:rl}b,c, it is negative, although small in the absolute value. On the other hand, the LE calculated on any periodic orbit of the Anosov map vanishes (because the M\"obius map is conjugated to a circle rotation). Moreover, it is well-known that periodic orbits in the Anosov map are dense. Thus, one cannot obtain the LE for a typical chaotic trajectory in the skew map \eqref{eq:bm} by virtue of the averaging over values corresponding to periodic orbits. The reason for this paradoxal situation is that the Lyapunov exponent in $x$-direction is not an ``observable'' along a trajectory of the Anosov map, but requires additional averaging over the dynamics of $x$ (and this averaging is not ensured when a chaotic trajectory ``jumps'' from one periodic orbit to another one).
\item \textbf{Moderate $|c|$:} In these transitional regions $c_1\leq c\leq c_2$ and $c_3\leq c\leq c_4$ some pairs of saddle periodic orbits already disappeared via a saddle-node bifurcation, but some other still exist. Attractor and repeller overlap, but their measures are concentrate in different regions (cf. Fig.~\ref{fig:ar}b). In Section~\ref{sec:hc}, we construct a heterodimensional cycle in the region $c_1\leq c\leq c_2$, i.e., a heteroclinic cycle connecting saddle periodic orbits inherited from the former attractor and the repeller. These periodic orbits have different dimensions of stable and unstable manifolds: those which belong to the attractor have two stable directions (one of them in $x$ variable) and one unstable, and those which belong to the repeller have two unstable directions (one of them in $x$ variable), and one stable. Thus, the dynamics in this region is heterodimensional.
\end{enumerate}

\subsection{First tangent bifurcation}
\label{sec:arc}

\begin{figure*}
\includegraphics[width=1\columnwidth]{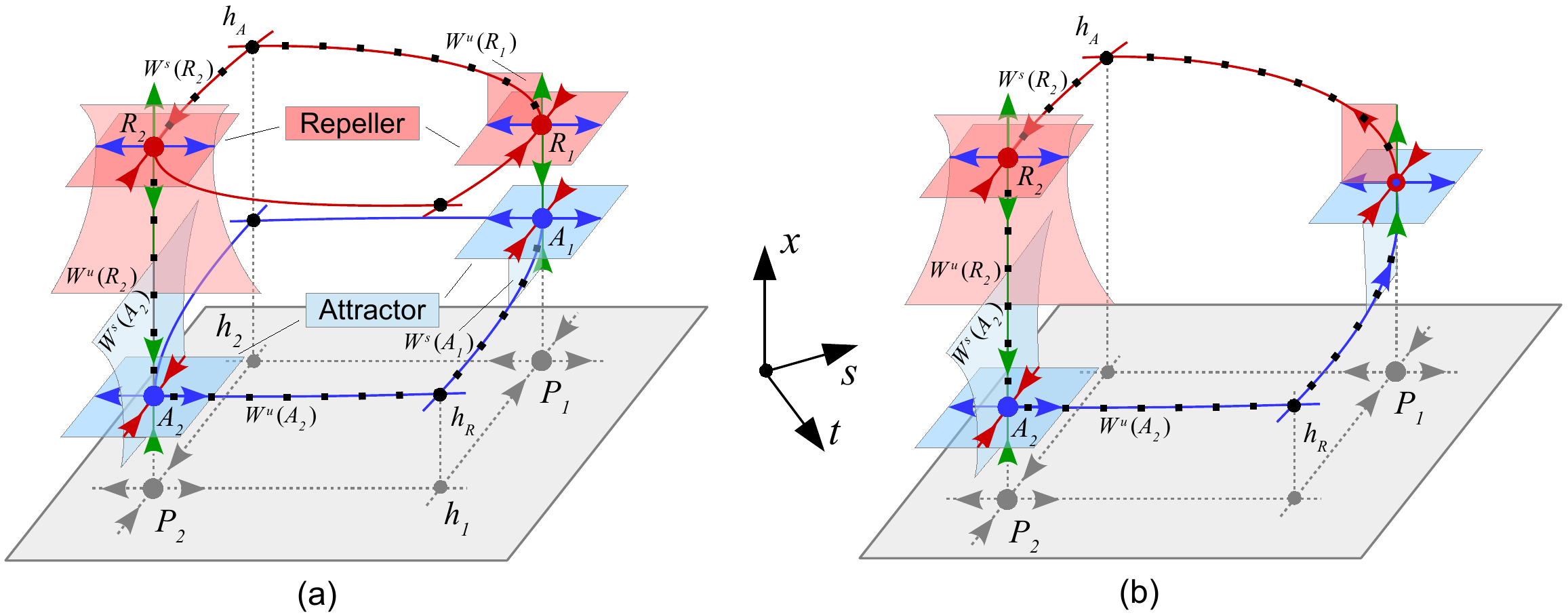}
\caption{Illustration of the attractor-repeller collision. (a) Separated chaotic attractor and chaotic repeller at $c_4(\alpha,\mu,\e) < c < c_1(\alpha,\mu,\e)$. $A_1$, $A_2$ and $R_1$, $R_2$ are fixed and period-2 saddle points belonging to the attractor and to the repeller, respectively (we show here every second iteration of the map, thus, both the fixed point and the period-2 point of the original system look like fixed points). (b) On the curves $c = c_1(\alpha,\mu,\e)$ and $c = c_4(\alpha,\mu,\e)$ one of pairs of periodic orbits, say $A_1$ and $R_1$, merges due to a tangent bifurcation (this situation is depicted in panel (b)), and disappears beyond it. Beyond the first such bifurcation, the attractor and repeller are no longer separated.}
\label{fig:AR_collision_sketch}
\end{figure*}

The aim of this section is to show how the attractor and repeller collide at the first tangent bifurcation, and what happens to heteroclinic connections. In region 1 each period-$m$ point $P^i_m$ in the Anosov subsystem \eqref{eq:bma1}--\eqref{eq:bma2} corresponds to a pair of period-$m$ orbits $A^i_m$ and $R^i_m$: $A^i_m$ belongs to the attractor, while $R^i_m$ belongs to the repeller. At the curves $c = c_1(\alpha,\mu,\e)$ and $c = c_4(\alpha,\mu,\e)$ one of pairs of these periodic orbits merges via a tangent bifurcation.
We illustrate this situation in Fig.~\ref{fig:AR_collision_sketch}.

For simplicity, we consider the case where the first tangent bifurcation occurs with fixed points $A_1$ and $R_1$. Such a bifurcation occurs as the first one when $c$ increases while parameter $\alpha$ is close to $\alpha = \pi/2$ in Fig.~\ref{fig:pbd}a. Below the curve $c = c_1(\alpha,\mu,\e)$, all periodic orbits $A^i_m$ of the attractor are intertwined in one homoclinic tangle formed by stable $W^s(A^i_m)$ and unstable $W^u(A^i_m)$ invariant manifolds (the same for the repeller). This means that there are heteroclinic connections between all periodic orbits of the attractor (the same for the repeller). Schematically, this is illustrated in Fig.~\ref{fig:AR_collision_sketch}a, where it is shown how the points $A_1$ and $A_2$ (and also $R_1$ and $R_2$) are connected via a heteroclinic cycle. The points $A_1$ and $R_1$ merge at a tangent bifurcation (this is a codimension-one bifurcation of a saddle-saddle fixed point possessing a homoclinic orbit~\cite{afraimovich1982bifurcation}) on the curve $c = c_1(\alpha,\mu,\e)$, see Fig.~\ref{fig:AR_collision_sketch}b, and disappear above it. One can see that in region 3, beyond the first tangent bifurcation, trajectories going from the former attractor to the former repeller become possible. We will further explore this in Section \ref{sec:hc}.

\subsection{Eyelet intermittency}

\begin{figure}
\includegraphics[width=0.6\columnwidth]{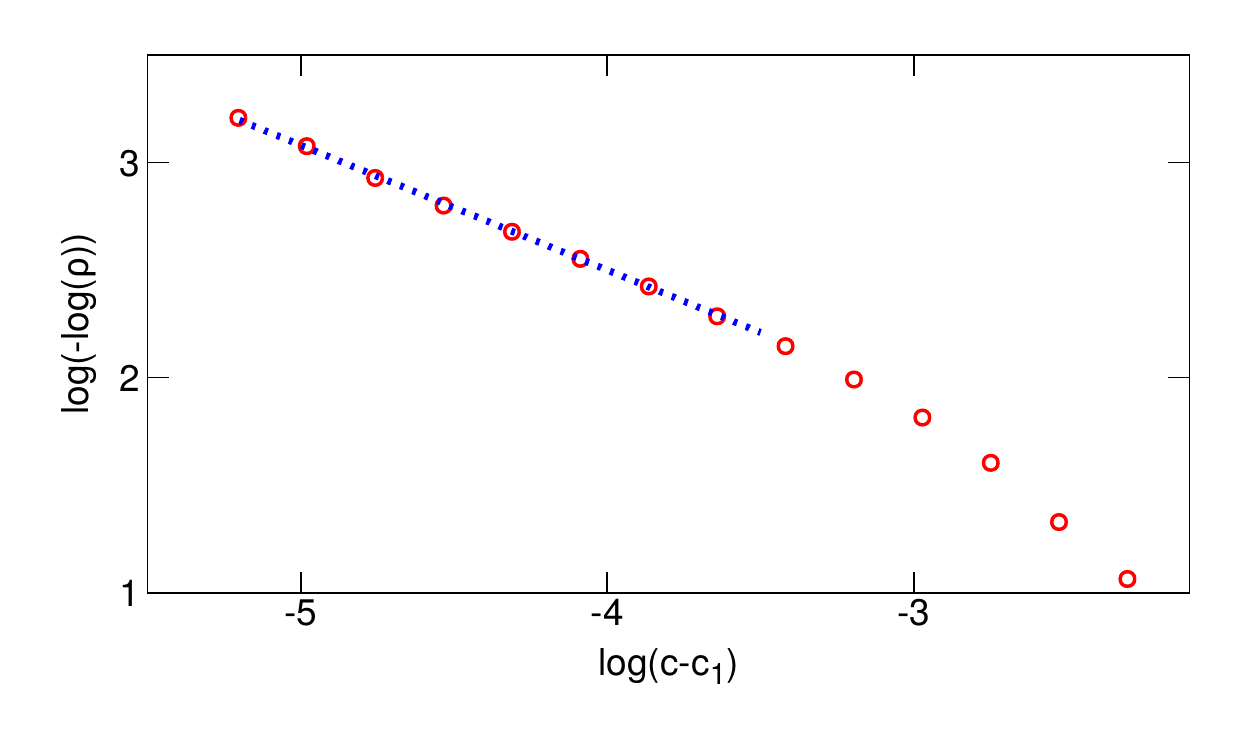}
\caption{Dependence
of the rotation number on the parameter $c$ close to criticality. Other parameters: $\e=0.4$, $\alpha=1.5$, $\mu=0.1$, $c_1= \arcsin(\e)/\pi-\mu\sin\alpha$. For the calculation
of the rotation number a trajectory of length $5\cdot 10^9$ was explored. The dotted straight line has a slope $-0.58$, which slightly
deviates from the theoretical prediction $0.5$.}
\label{fig:eye}
\end{figure}

The transition from regime 1 to regime 3, described in Section \ref{sec:arc},
often called attractor-repeller collision, has been studied in papers~\cite{Grebogi-Ott-Yorke-83c, Grebogi-Ott-Yorke-85, Pikovsky-Osipov-Rosenblum-Zaks-Kurths-97}. The basic observation is that at the initial stage of this transition (i.e., where the regime 3 has just appeared), the rotation number $\rho$ grows very slowly~\cite{Grebogi-Ott-Yorke-83c, Grebogi-Ott-Yorke-85, Pikovsky-Osipov-Rosenblum-Zaks-Kurths-97}, namely as $\log (\rho)\sim -\Delta^{-1/2}$, where $\Delta$ is the deviation of parameter $c$ from it critical value (in the situation above $\Delta=c-c_{1}$). We illustrate this statistical law in Fig.~\ref{fig:eye}, where we show the rotation number in a double logarithmic scale. The reason for such a slow growth of the rotation number lies in the transversal instability of the trajectory connecting former attractor and repeller beyond the first tangent bifurcation at $c=c_1$. Indeed, for a circle map beyond the tangent bifurcation, the characteristic time $\tau$ to perform a full rotation scales as $\tau\sim \Delta^{-1/2}$. Thus, in the full system one can observe such a rotation only if a chaotic trajectory spends at least this time in a small vicinity (say, of size $\delta$) of the saddle periodic orbit in the driving system \eqref{eq:bma1}--\eqref{eq:bma2}, which is responsible for the tangent bifurcation. Because this periodic orbit is a saddle with the unstable Lyapunov exponent $\lambda^u$, an initial point should be in the vicinity of size $\delta e^{-\lambda^u \tau}$ of the periodic orbit in order to remain in this vicinity for time $\tau$. The probability for this is $\sim \delta e^{-\lambda^u \tau}$, thus one needs $T\sim \delta^{-1} e^{\lambda^u \tau}$ iterations of the original system to perform one rotation. This yields the estimation of the rotation number $\rho\sim T^{-1}\sim \delta  e^{-\lambda^u \Delta^{-1/2}}$, from which the expression for the rotation number near the criticality follows (see also Ref.~\onlinecite{olicon2020critical} for a similar expression for
a circle map with noisy, not chaotic, driving). Although a trajectory connecting attractor and repeller is rather unprobable close to the transition point $c_{1}$ in a ``free run'', such a trajectory can be explicitly constructed, as we will discuss in Section \ref{sec:hc}.

We conclude this section with a discussion of validity of the approach to the attractor-repeller collision by virtue of following bifurcations of periodic orbits embedded in chaos. We are not aware of rigorous results proving that indeed such a collision is mediated by the periodic orbit which first undergoes saddle-node bifurcation. However, the diagram shown in Fig.~\ref{fig:pbd} indicates that there are parameter regions where cycles with low periods bifurcate definitely prior to cycles with higher periods. Thus one hardly expects that a non-periodic orbit can bifurcate prior to the low-periodic cycles. On the other hand, such a case cannot be excluded in general. A search for a situation where a chaotic transition is mediated by a non-periodic orbit appears to be a challenging timely problem in the theory of chaotic transitions.

\section{Heterodimensional cycles}
\label{sec:hc}

In this section we study the intermediate situation (type 3 above) in more detail.
Here the attractor and the repeller overlap (this means that at least one pair of
periodic orbits already disappeared via a tangent bifurcation), and this topological
object possesses still many saddle periodic orbits from both former attractor and repeller.
Let us call them for brevity \textit{A-orbits} and \textit{R-orbits}, respectively. These saddle orbits have different dimensions of stable and unstable manifolds: A-orbits have a two-dimensional stable and one-dimensional unstable manifolds (one stable and one unstable directions from the Anosov map \eqref{eq:bma1}--\eqref{eq:bma1}, and one stable eigenvector in the $x$-direction), R-orbits have a one-dimensional stable and two-dimensional unstable manifolds (here, the eigenvector in the $x$-direction is unstable). This situation is sometimes called \textit{unstable dimension variability} (UDV)~\cite{KOSTELICH199781, Das-Yorke-17, alligood2006crossing}, it has been discussed in mathematical literature~\cite{AbrahamSmale70, diaz1995robust, Li_2017, Li_17, hittmeyer2018existence, Li_2020, hittmeyer2020identify, Hammerlindl_22} and in applications~\cite{Kuptsov_2013, Vallejo-Sanjuan-16, Esashi_etal-18}.

\begin{figure*}
\includegraphics[width=1\columnwidth]{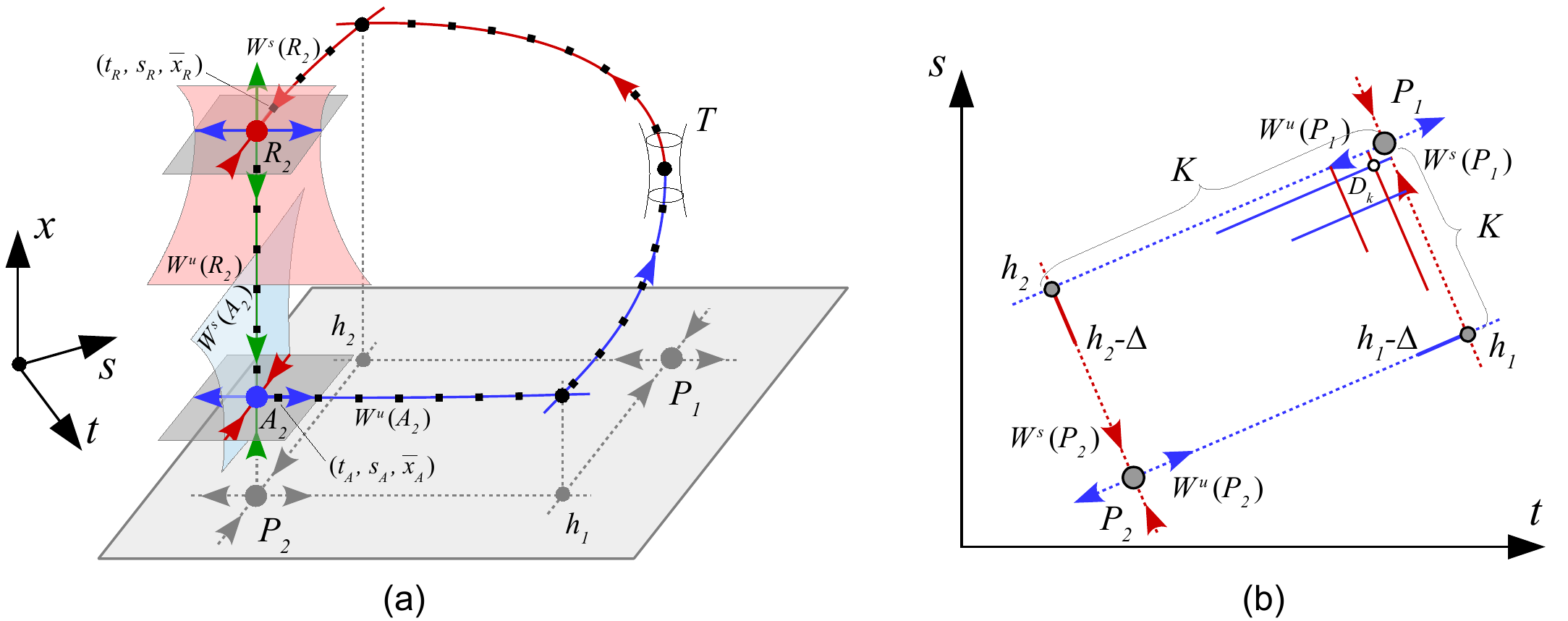}
\caption{(a) Illustration for the numerical construction of a heterodimensional cycle connecting the points $A_2$ and $R_2$. Despite the points $A_2$ and $R_2$ this cycle also consists of two orbits (marked by black square dots). One (trivial) orbit belongs to the transversal intersection of two-dimensional manifolds $W^s(A_2)$ and $W^u(R_2)$,
along this orbit only variable $x$ varies (so it looks like a vertical line in the figure). Another (nontrivial) orbit passes through the non-transversal (codimension-1) intersection of one-dimensional manifolds $W^u(A_2)$ and $W^s(R_2)$ inside a narrow tunnel $T$. The points $P_1$ and $P_2$ are the fixed and the period-2 points in the Anosov map \eqref{eq:bma1}--\eqref{eq:bma2}; the heteroclicnic points $h_1$ and $h_2$ belong to the intersection $W^u(P_2) \cap W^s(P_1)$ and $W^s(P_2) \cap W^u(P_1)$, respectively. (b) Heteroclinic cycle connecting the points $P_1$ and $P_2$. This construction is used to find a homoclinic orbit to the point $P_2$ in the Anosov map \eqref{eq:bma1}--\eqref{eq:bma2}. This homoclinic orbit of the Anosov map is then used as a driving trajectory for numerical construction of the heterodimensional cycle.}
\label{fig:sketch}
\end{figure*}

A characteristic feature of this regime is the existence of heterodimensional cycles \cite{BonattiDiaz96, BonattiDiaz08, LiTuraev2021} consisting of pairs of heteroclinic trajectories connecting A-orbits with R-orbits: one of these trajectories lies in the transversal intersection of two-dimensional manifolds of A- and R-orbits, while another one passes through a non-transversal (codimension-1) intersection of one-dimensional manifolds of these orbits. The main goal of this section is to provide a numerical evidence for the existence of such cycles in the intermediate case 3. For simplicity, we restrict ourselves to the simplest case where the A- and R-orbits have period two, while the pair of fixed points $A_1$ and $R_1$ which used to belong to the attractor and repeller, respectively, already disappeared via a tangent bifurcation (cf. Fig.~\ref{fig:AR_collision_sketch}).

In this case, it is more convenient to consider the second iteration of the map. Then, the period-2 points $A_2$ and $R_2$ in map \eqref{eq:bm} becomes fixed points. The corresponding fixed point of the twice-iterated Anosov map \eqref{eq:bma1}--\eqref{eq:bma2} is denoted $P_2$. The fixed point of the Anosov map is denoted $P_1$. The cycle that we will construct, starts at $A_2$. Then, the $(t,s)$-values become close to the fixed point $P_1$ and, at these iterations, $x$-values move from the former attractor to the former repeller through the narrow tunnel (region T in Fig.~\ref{fig:sketch}a) located on the place where the fixed points $A_1$ and $R_1$ existed before the tangent bifurcations. After this, the trajectory comes asymptotically close to $R_2$. We sketch this trajectory in Fig.~\ref{fig:sketch}a.

For the numerical construction of the cycle described above we choose the following values of parameters:
$
\mu = 0.8,\; \varepsilon = 0.4,\; \alpha = 1.5
$.
Indeed, for these values of the parameter the M\"obius map \eqref{eq:bmm} at the point $P_1$ is conjugated to a rotation (i.e., parameters are taken beyond the tangent bifurcation of the corresponding fixed points $A_1$ and $R_1$), but the $A_2$- and $R_2$-points still exist. In Fig.~\ref{fig:pbd} this situation occurs near $\alpha\approx \pi/2$ above the blue curve corresponding to the tangent bifurcation of the fixed points $A_1$ and $R_1$ but below other curves corresponding to the tangent bifurcations of periodic points of higher periods (the adopted parameter value $\alpha=1.5$ is marked with a dashed cyan line in Fig.~\ref{fig:pbd}).

\begin{figure}
\includegraphics[width=0.7\columnwidth]{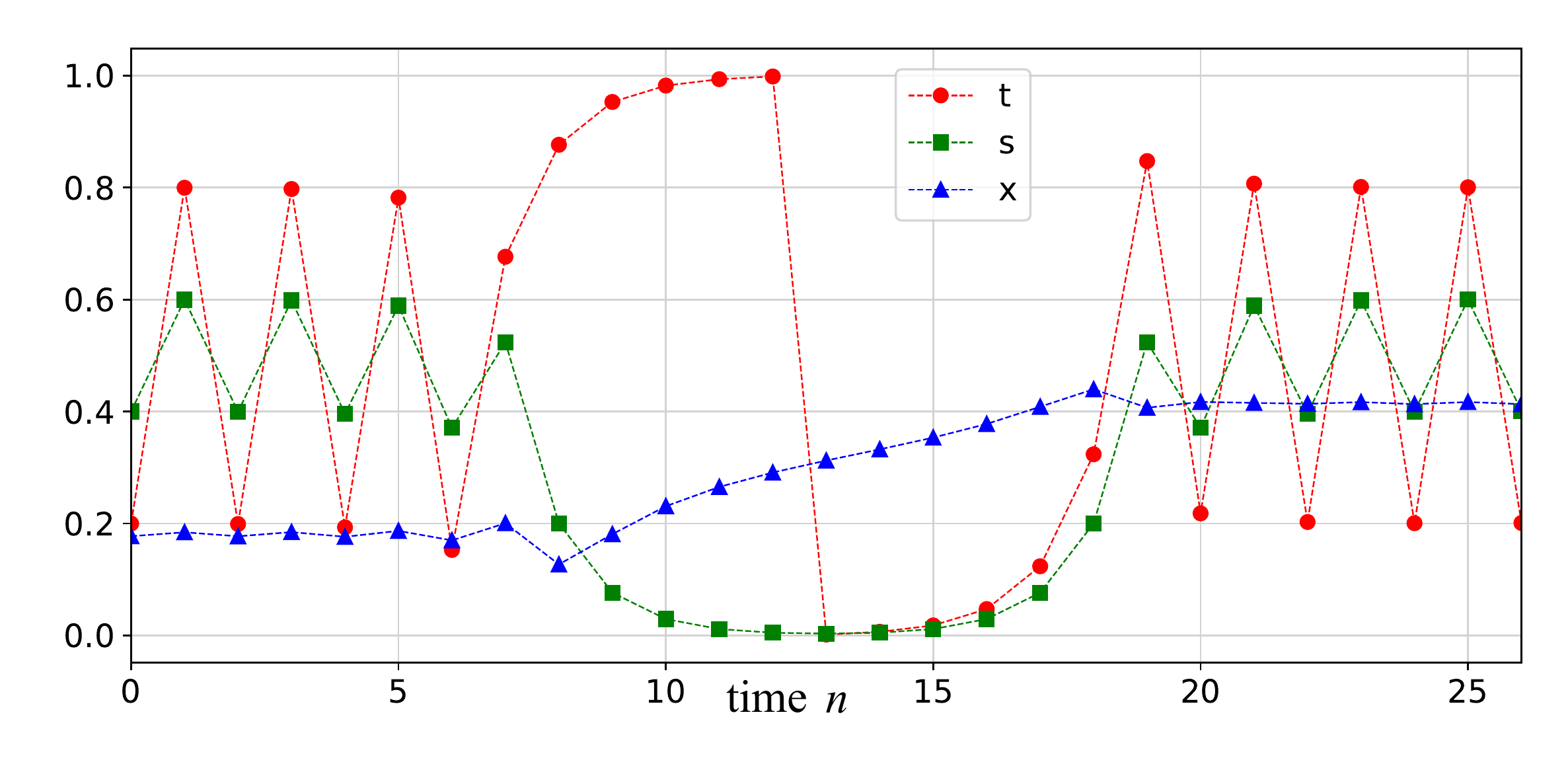}
\caption{Illustration of the constructed heteroclinic orbit. Parameters: $\e=0.4$, $\alpha=1.5$, $\mu=0.08$, $c=0.07112495671202002$. One can see that in variables $t,s$ (red and green markers) this trajectory is a homoclinic trajectory to the period-2 cycle (epochs $n<5$ and $n>20$), which passes close to the fixed point $t=s=0$ (epoch $10\lesssim n\lesssim 15$). During the stay of the $(t,s)$ trajectory close to the fixed point, variable $x$ (blue markers) varies from $x\approx 0.2$ (position of the attractor) to $x\approx 0.4$ (position of the repeller).}
\label{fig:horbit}
\end{figure}

The strategy for this numerical construction consists of two stages. First, we compute a driving one trajectory in the Anosov map \eqref{eq:bma1}--\eqref{eq:bma2} as a homoclinic trajectory for the point $P_2$ which comes very close to the fixed point $P_1$. At the next stage, we use this trajectory as the driving in the M\"obius map \eqref{eq:bmm} in order to construct the full heteroclinic cycle connecting points $A_2$ and $R_2$.
\begin{itemize}
\item First, on the $(t,s)$-plane we find intersection points $h_1$ and $h_2$ of the unstable manifold $W^u(P_2)$ with the stable manifold $W^s(P_1)$, and of the unstable manifold $W^u(P_1)$ with the stable manifold $W^s(P_2)$, respectively, see Fig.~\ref{fig:sketch}b. This is a straightforward task, because all the manifolds are straight lines. Thus, we construct two heteroclinic connections $P_2 \to P_1$ and $P_1 \to P_2$.
\item Then, we find a homoclinic trajectory of the Anosov map \eqref{eq:bma1}--\eqref{eq:bma2} $P_2\to P_2$ which passes close to the constructed heteroclinic cycles. For this goal, on the unstable manifold $W^u(P_2)$, we take a small segment $[h_1-\Delta, h_1]$ where the point $h_1-\Delta$ lies between the points $P_2$ and $h_1$. We iterate this segment forward in time (say, we use $K$ iterations), until the iteration of the point $h_1$ comes sufficiently close to the point $P_1$. Similarly, on the stable manifold $W^s(P_2)$ we take a small segment $[h_2-\Delta, h_2]$ where the point $h_2-\Delta$ lies between the points $P_2$ and $h_2$. We iterate it backward in time (again $K$ times) until the iteration of $h_2$ comes close to the point $P_1$. This ensures that the corresponding images of the segments $[h_1-\Delta, h_1]$ and $[h_2-\Delta, h_2]$ intersect at some point $D_k$, which is as close to the point $P_1$ as we want (increasing $K$ we can impose $D_k$ to be arbitrarily close to $P_1$).
\item Iterations of the point $D_k$ in the Anosov map \eqref{eq:bma1}--\eqref{eq:bma2} give a homoclinic orbit to the point $P_2$, which comes very close to the fixed point $P_1$ and, thus, spends a large time in the vicinity of this point. We will use it as a driving force for the M\"obius map \eqref{eq:bmm}.
\item In the next step we find a point $(t_A,s_A,x_A)$ on the unstable manifold $W^u(A_2)$ very close to the point $A_2$. For this, we take a point $(t_A,s_A)$ which is very close to the point $P_2$ and, at the same time, belongs to the trajectory of $D_k$. Then, we select several values $x_i$ close to the $x$-coordinate of the point $A_2$, and iterate the points $(t_A,s_A,x_i)$ backward in time. The $(t,s)$-values converge to the point $P_2$, while values of $x$ either grow or decrease except for those that belong to $W^u(A_2)$. Taking more $x$-values between the neighboring points which, by these backward iterations, go along different branches (up and down) of the stable manifold $W^s(A_2)$ in the $x$-direction, we can find a point $(t_A,s_A,x_A)$ on the unstable manifold $W^u(A_2)$ with desired accuracy. In the same way we find a point $(x_R,t_R,s_R)$ lying on the one-dimensional stable manifold of the point $R_2$.
\item Finally, we iterate the points $(t_A,s_A,x_A)$ and $(x_R,t_R,s_R)$, respectively, forward and backward in time until their $(t,s)$-coordinates reach the point $D_k$. Generally at this point the resulting $x$-coordinates $\overline{x}_A$ and $\overline{x}_R$ do not coincide. However, varying one of the parameters in map \eqref{eq:bm} (we have varied parameter $c$), we can find a value at which $\overline{x}_A=\overline{x}_B$. This completes the construction of heteroclinic connection between one dimensional manifolds $W^u(A_2)$ and $W^s(R_2)$ of a particular heterodimensional cycle. The constructed trajectory is illustrated in Fig.~\ref{fig:horbit}.
\item Note that the intersection of two-dimensional manifolds $W^u(R_2)$ and $W^s(A_2)$ providing the second heteroclinic connection between the points $A_1$ and $A_2$ exists always. Thus, the described procedure gives a numerical evidence for the existence of heterodimensional cycles.
\end{itemize}

In fact, according to the described above procedure, many heterodimensional cycles could be constructed, because there are (infinitely) many different homoclinic orbits to the point $P_2$, passing close to the point $P_1$ in the Anosov map. This is, however, not needed, because according to theory developed in Ref.~\onlinecite{LiTuraev2021}, the existence of a (general) heterodimensional cycle implies the existence of many such cycles in a vicinity of parameters values. Our calculations, thus, confirm that at the attractor-repeller collision described by map \eqref{eq:bm}, the heterodimensional dynamics appear. Remarkably, such a regime disappears when the ``last'' periodic orbits on the attractor and the repeller disappear via a tangent bifurcation (i.e. the system enters regime 2 in the
classification of Section~\ref{sec:bpo}). This is a particularity of the M\"obius map, which has only fixed points but not isolated periodic orbits of higher periods.

\section{Variability of the finite-time Lyapunov exponents}
\label{sec:var}

\begin{figure}
\includegraphics[width=0.7\columnwidth]{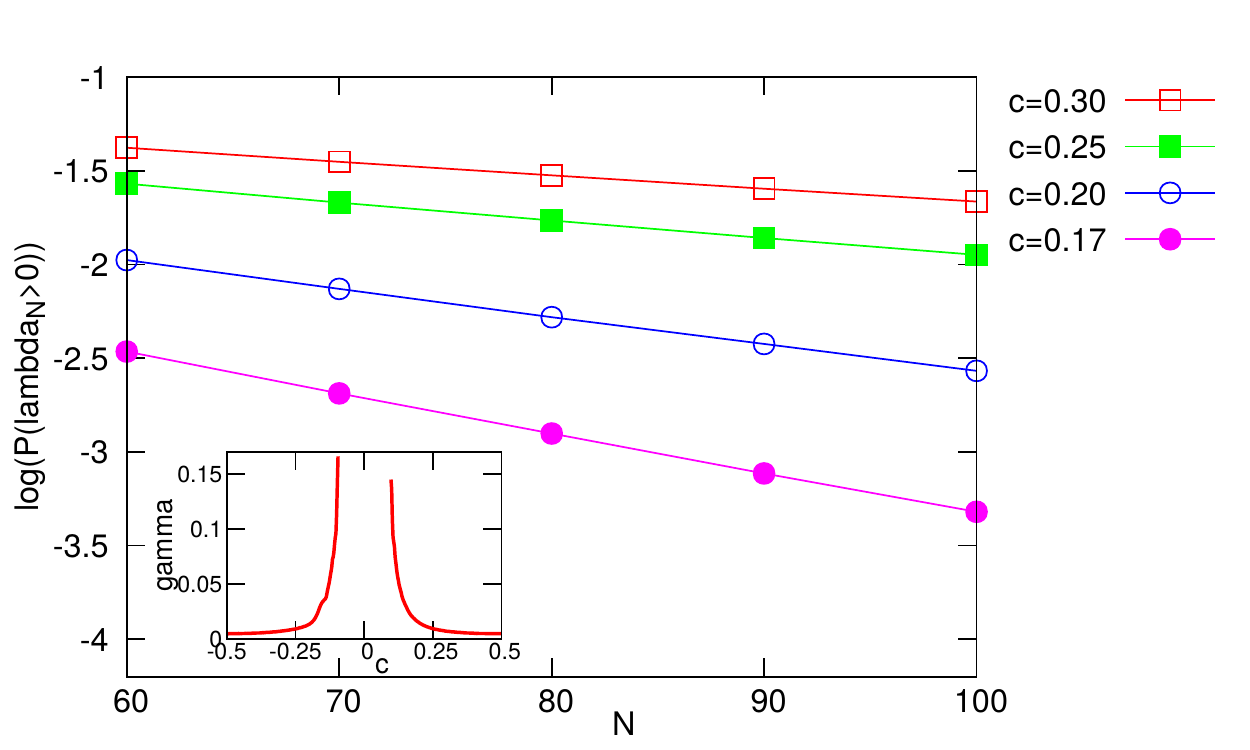}
\caption{Illustration of the scaling $P(\lambda_N>0)\sim \exp[-N \gamma]$ for several
values of parameter $c$ for $\e=0.4$, $\mu=0.1$, $\alpha=1.5$. Markers: calculated probabilities; lines: exponential fit of the data points.
The inset shows the
exponent $\gamma$  as a function of parameter $c$.
For small values of $c$, where the attractor and the repeller are separated, this exponent is not defined because the finite-time exponents are all negative for large enough $N$.}
\label{fig:ftle}
\end{figure}

One of the implications of the heterodimensional dynamics is a strong variability of the Lyapunov exponent. Indeed, consider a situation where the attractor and the repeller overlap and they possess periodic orbits with different stability in the $x$-direction. Then a trajectory which visits vicinities of these periodic orbits will have positive and negative finite-time Lyapunov exponents, depending on whether this trajectory is close to unstable or to stable in $x$-direction periodic orbits, respectively. This means that the spread of the finite-time Lyapunov exponents is large and includes positive values, although the average Lyapunov exponent is negative.

To test this property of the heterodimensional dynamics in our system, we calculated finite-time Lyapunov exponents $\lambda_N=\frac{1}{N}\sum_{k=n}^{k=n+N-1} \log \frac{d x_{k+1}}{d x_k}$ for time intervals up to $N=100$. According to a general scaling of finite-time Lyapunov exponents~\cite{Pikovsky-Politi-16}, the probability to have positive values scales as $P(\lambda_N>0)\sim \exp[-N \gamma]$ with some positive exponent $\gamma$. We illustrate this scaling in Fig.~\ref{fig:ftle} for several values
of parameter $c$. At this point we would like to justify the chosen range of the time intervals $N$. According to the theory, one should take $N$ as large as possible for the asymptotic exponential decay of probability above to hold. On the other hand, reliable estimation of very small probabilities is computationally challenging, because it requires very long time series. Figure~\ref{fig:ftle} shows that the chosen range of values of $N$ appears reasonable, because in this full range the dependence of the probability on $N$ appears as exponential with a good precision. Furthermore, we show in the inset the
dependence of this exponent on parameter $c$. One can see that close to the attractor-repeller collision this exponent is large, corresponding to a very small probability to visit the domain of the former repeller. This exponent is small in the domain where the invariant measures of the attractor and the repeller nearly coincide (values of $|c|$ close to $0.5$).

\section{Conclusion}
\label{sec:concl}

The main goal of this paper was to show that robust heterodimensional dynamics naturally appears beyond the attractor-repeller collision in a chaotically driven circle map. We have considered a particular model where the driving system is a hyperbolic Anosov torus map, and the circle map is a M\"obius map. Since both subsystems are easily invertible, the attractor and the repeller can simply be followed as the limiting sets of forward and backward iterations, respectively. We traced the collision via tangent bifurcations of the periodic orbits embedded in chaos. Furthermore, we constructed nontrivial heterodimensional cycles inside the overlapped attractor and repeller as trajectories, connecting remaining, not yet bifurcated periodic orbits from the former attractor and repeller. This allows the application of a general theory about the persistence of heterodimensional dynamics~\cite{LiTuraev2021}. The demonstrated relation of the attractor-repeller collision
to the concept of heterodimensional dynamics is the main contribution of this study compared to previous explorations of this transition~\cite{Grebogi-Ott-Yorke-83c, Grebogi-Ott-Yorke-85,Pikovsky-Osipov-Rosenblum-Zaks-Kurths-97}.

We also explored properties of the Lyapunov exponent in the driven circle map. This exponent is always negative, which is not surprising for separated attractors and repellers and for a situation close to the attractor-repeller collision. However, we demonstrate that the Lyapunov exponent is also negative, although small in absolute value, in a situation where all periodic orbits in the system have disappeared via a tangent bifurcation. Here, due to peculiar properties of the M\"obius map, on every periodic orbit of the driving Anosov map the Lyapunov exponent of the driven map vanishes. Nevertheless, on average the Lyapunov exponent remains negative. Another interesting property of the Lyapunov exponent is that in the case of heterodimensional dynamics, finite-time Lyapunov exponents in the $x$-direction can be positive. This is because the unstable (in the direction of the circle map)
periodic orbits belong to the chaotic set.
We confirm that the probability of observing a positive time-$N$ exponent
decreases exponentially with $N$.

\acknowledgments
We thank Dmitry~Turaev, for fruitful discussions.
This paper was supported by the RSF grant No. 19-71-10048 (Sections \ref{sec:bm}, \ref{sec:parc}, and \ref{sec:hc}). Sections \ref{sec:cupo}, \ref{sec:var}, and \ref{sec:concl} were prepared within the framework of the Basic Research Program at HSE University. 

\section*{Data availability}
All numerical experiments with two-dimensional maps are described in the paper and can be reproduced without additional information.


\begin{thebibliography}{10}%
\makeatletter
\providecommand \@ifxundefined [1]{%
 \ifx #1\undefined \expandafter \@firstoftwo
 \else \expandafter \@secondoftwo
\fi
}%
\providecommand \@ifnum [1]{%
 \ifnum #1\expandafter \@firstoftwo
 \else \expandafter \@secondoftwo
\fi
}%
\providecommand \enquote [1]{``#1''}%
\providecommand \bibnamefont  [1]{#1}%
\providecommand \bibfnamefont [1]{#1}%
\providecommand \citenamefont [1]{#1}%
\providecommand\href[0]{\@sanitize\@href}%
\providecommand\@href[1]{\endgroup\@@startlink{#1}\endgroup\@@href}%
\providecommand\@@href[1]{#1\@@endlink}%
\providecommand \@sanitize [0]{\begingroup\catcode`\&12\catcode`\#12\relax}%
\@ifxundefined \pdfoutput {\@firstoftwo}{%
 \@ifnum{\z@=\pdfoutput}{\@firstoftwo}{\@secondoftwo}%
}{%
 \providecommand\@@startlink[1]{\leavevmode}%
 \providecommand\@@endlink[0]{}%
}{%
 \providecommand\@@startlink[1]{%
  \leavevmode
  \pdfstartlink
   attr{/Border[0 0 1 ]/H/I/C[0 1 1]}%
   user{/Subtype/Link/A<</Type/Action/S/URI/URI(#1)>>}%
  \relax
 }%
 \providecommand\@@endlink[0]{\pdfendlink}%
}%
\providecommand \url  [0]{\begingroup\@sanitize \@url }%
\providecommand \@url [1]{\endgroup\@href {#1}{\urlprefix}}%
\providecommand \urlprefix [0]{URL }%
\providecommand \Eprint[0]{\href }%
\@ifxundefined \urlstyle {%
  \providecommand \doi [1]{doi:\discretionary{}{}{}#1}%
}{%
  \providecommand \doi [0]{doi:\discretionary{}{}{}\begingroup
  \urlstyle{rm}\Url }%
}%
\providecommand \doibase [0]{http://dx.doi.org/}%
\providecommand \Doi[1]{\href{\doibase#1}}%
\providecommand \selectlanguage [0]{\@gobble}%
\providecommand \bibinfo [0]{\@secondoftwo}%
\providecommand \bibfield [0]{\@secondoftwo}%
\providecommand \translation [1]{[#1]}%
\providecommand \BibitemOpen[0]{}%
\providecommand \bibitemStop [0]{}%
\providecommand \bibitemNoStop [0]{.\EOS\space}%
\providecommand \EOS [0]{\spacefactor3000\relax}%
\providecommand \BibitemShut [1]{\csname bibitem#1\endcsname}%
\bibitem{BonattiDiaz96}%
  \BibitemOpen
  \bibfield{author}{%
  \bibinfo {author} {\bibfnamefont{C.}~\bibnamefont{Bonatti}}\ and\ \bibinfo
  {author} {\bibfnamefont{L.~J.}\ \bibnamefont{D{\'i}az}},\ }%
  \bibfield{title}{%
  \enquote{\bibinfo {title} {Persistent nonhyperbolic transitive
  diffeomorphisms},}\ }%
  \bibfield{journal}{%
  \bibinfo {journal} {Annals of Mathematics},\ \bibinfo {pages} {357--396}}%
   (\bibinfo {year} {1996})\BibitemShut{NoStop}%
\bibitem{AbrahamSmale70}%
  \BibitemOpen
  \bibfield{author}{%
  \bibinfo {author} {\bibfnamefont{R.}~\bibnamefont{Abraham}}\ and\ \bibinfo
  {author} {\bibfnamefont{S.}~\bibnamefont{Smale}},\ }%
  \bibfield{title}{%
  \enquote{\bibinfo {title} {Nongenericity of {$\Omega$}-stability},}\ }%
  \bibfield{journal}{%
  \bibinfo {journal} {Math. Soc., Providence, R.I.}\ }%
  \textbf{\bibinfo {volume} {14}},\ \bibinfo {pages} {5--8} (\bibinfo {year}
  {1970})\BibitemShut{NoStop}%
\bibitem{DiazRocha92}%
  \BibitemOpen
  \bibfield{author}{%
  \bibinfo {author} {\bibfnamefont{L.~J.}\ \bibnamefont{D{\'\i}az}}\ and\
  \bibinfo {author} {\bibfnamefont{J.}~\bibnamefont{Rocha}},\ }%
  \bibfield{title}{%
  \enquote{\bibinfo {title} {Nonconnected heterodimensional cycles: bifurcation
  and stability},}\ }%
  \bibfield{journal}{%
  \bibinfo {journal} {Nonlinearity}\ }%
  \textbf{\bibinfo {volume} {5}},\ \bibinfo {pages} {1315} (\bibinfo {year}
  {1992})\BibitemShut{NoStop}%
\bibitem{diaz1995robust}%
  \BibitemOpen
  \bibfield{author}{%
  \bibinfo {author} {\bibfnamefont{L.~J.}\ \bibnamefont{D{\'\i}az}},\ }%
  \bibfield{title}{%
  \enquote{\bibinfo {title} {Robust nonhyperbolic dynamics and
  heterodimensional cycles},}\ }%
  \bibfield{journal}{%
  \bibinfo {journal} {Ergodic Theory and Dynamical Systems}\ }%
  \textbf{\bibinfo {volume} {15}},\ \bibinfo {pages} {291--315} (\bibinfo
  {year} {1995})\BibitemShut{NoStop}%
\bibitem{diaz1995persistence}%
  \BibitemOpen
  \bibfield{author}{%
  \bibinfo {author} {\bibfnamefont{L.~J.}\ \bibnamefont{D{\'\i}az}},\ }%
  \bibfield{title}{%
  \enquote{\bibinfo {title} {Persistence of cycles and nonhyperbolic dynamics
  at heteroclinic bifurcations},}\ }%
  \bibfield{journal}{%
  \bibinfo {journal} {Nonlinearity}\ }%
  \textbf{\bibinfo {volume} {8}},\ \bibinfo {pages} {693} (\bibinfo {year}
  {1995})\BibitemShut{NoStop}%
\bibitem{BonattiDiaz08}%
  \BibitemOpen
  \bibfield{author}{%
  \bibinfo {author} {\bibfnamefont{C.}~\bibnamefont{Bonatti}}\ and\ \bibinfo
  {author} {\bibfnamefont{L.~J.}\ \bibnamefont{D{\'\i}az}},\ }%
  \bibfield{title}{%
  \enquote{\bibinfo {title} {Robust heterodimensional cycles and-generic
  dynamics},}\ }%
  \bibfield{journal}{%
  \bibinfo {journal} {Journal of the Institute of Mathematics of Jussieu}\ }%
  \textbf{\bibinfo {volume} {7}},\ \bibinfo {pages} {469--525} (\bibinfo {year}
  {2008})\BibitemShut{NoStop}%
\bibitem{LiTuraev2021}%
  \BibitemOpen
  \bibfield{author}{%
  \bibinfo {author} {\bibfnamefont{D.}~\bibnamefont{Li}}\ and\ \bibinfo
  {author} {\bibfnamefont{D.}~\bibnamefont{Turaev}},\ }%
  \bibfield{title}{%
  \enquote{\bibinfo {title} {Persistence of heterodimensional cycles},}\ }%
  \bibfield{journal}{%
  \bibinfo {journal} {arXiv preprint arXiv:2105.03739}}%
   (\bibinfo {year} {2021})\BibitemShut{NoStop}%
\bibitem{Pikovsky-Osipov-Rosenblum-Zaks-Kurths-97}%
  \BibitemOpen
  \bibfield{author}{%
  \bibinfo {author} {\bibfnamefont{A.}~\bibnamefont{Pikovsky}}, \bibinfo
  {author} {\bibfnamefont{G.~V.}\ \bibnamefont{Osipov}}, \bibinfo {author}
  {\bibfnamefont{M.}~\bibnamefont{Rosenblum}}, \bibinfo {author}
  {\bibfnamefont{M.}~\bibnamefont{Zaks}},\ and\ \bibinfo {author}
  {\bibfnamefont{J.}~\bibnamefont{Kurths}},\ }%
  \bibfield{title}{%
  \enquote{\bibinfo {title} {Attractor-repeller collision and eyelet
  intermittency at the transition to phase synchronization},}\ }%
  \bibfield{journal}{%
  \bibinfo {journal} {Phys. Rev. Lett.}\ }%
  \textbf{\bibinfo {volume} {79}},\ \bibinfo {pages} {47--50} (\bibinfo {year}
  {1997})\BibitemShut{NoStop}%
\bibitem{gonchenko2017phenomenon}%
  \BibitemOpen
  \bibfield{author}{%
  \bibinfo {author} {\bibfnamefont{A.~S.}\ \bibnamefont{Gonchenko}}, \bibinfo
  {author} {\bibfnamefont{S.~V.}\ \bibnamefont{Gonchenko}}, \bibinfo {author}
  {\bibfnamefont{A.}~\bibnamefont{Kazakov}},\ and\ \bibinfo {author}
  {\bibfnamefont{D.}~\bibnamefont{Turaev}},\ }%
  \bibfield{title}{%
  \enquote{\bibinfo {title} {On the phenomenon of mixed dynamics in
  {P}ikovsky--{T}opaj system of coupled rotators},}\ }%
  \bibfield{journal}{%
  \bibinfo {journal} {Physica D: Nonlinear Phenomena}\ }%
  \textbf{\bibinfo {volume} {350}},\ \bibinfo {pages} {45--57} (\bibinfo {year}
  {2017})\BibitemShut{NoStop}%
\bibitem{Kazakov2019}%
  \BibitemOpen
  \bibfield{author}{%
  \bibinfo {author} {\bibfnamefont{A.~O.}\ \bibnamefont{Kazakov}},\ }%
  \bibfield{title}{%
  \enquote{\bibinfo {title} {On the appearance of mixed dynamics as a result of
  collision of strange attractors and repellers in reversible systems},}\ }%
  \bibfield{journal}{%
  \bibinfo {journal} {Radiophysics and Quantum Electronics}\ }%
  \textbf{\bibinfo {volume} {61}},\ \bibinfo {pages} {650--658} (\bibinfo
  {year} {2019})\BibitemShut{NoStop}%
\bibitem{kazakov2020merger}%
  \BibitemOpen
  \bibfield{author}{%
  \bibinfo {author} {\bibfnamefont{A.}~\bibnamefont{Kazakov}},\ }%
  \bibfield{title}{%
  \enquote{\bibinfo {title} {Merger of a {H}{\'e}non-like attractor with a
  {H}{\'e}non-like repeller in a model of vortex dynamics},}\ }%
  \bibfield{journal}{%
  \bibinfo {journal} {Chaos: An Interdisciplinary Journal of Nonlinear
  Science}\ }%
  \textbf{\bibinfo {volume} {30}},\ \bibinfo {pages} {011105} (\bibinfo {year}
  {2020})\BibitemShut{NoStop}%
\bibitem{gonchenko2016reversible}%
  \BibitemOpen
  \bibfield{author}{%
  \bibinfo {author} {\bibfnamefont{S.~V.}\ \bibnamefont{Gonchenko}},\ }%
  \bibfield{title}{%
  \enquote{\bibinfo {title} {Reversible mixed dynamics: A concept and
  examples},}\ }%
  \bibfield{journal}{%
  \bibinfo {journal} {Discontinuity, Nonlinearity, and Complexity}\ }%
  \textbf{\bibinfo {volume} {5}},\ \bibinfo {pages} {365--374} (\bibinfo {year}
  {2016})\BibitemShut{NoStop}%
\bibitem{Gonchenko2017}%
  \BibitemOpen
  \bibfield{author}{%
  \bibinfo {author} {\bibfnamefont{S.~V.}\ \bibnamefont{Gonchenko}}\ and\
  \bibinfo {author} {\bibfnamefont{D.}~\bibnamefont{Turaev}},\ }%
  \bibfield{title}{%
  \enquote{\bibinfo {title} {On three types of dynamics and the notion of
  attractor},}\ }%
  \bibfield{journal}{%
  \bibinfo {journal} {Proceedings of the Steklov Institute of Mathematics}\ }%
  \textbf{\bibinfo {volume} {297}},\ \bibinfo {pages} {116--137} (\bibinfo
  {year} {2017})\BibitemShut{NoStop}%
\bibitem{turaev2021criterion}%
  \BibitemOpen
  \bibfield{author}{%
  \bibinfo {author} {\bibfnamefont{D.}~\bibnamefont{Turaev}},\ }%
  \bibfield{title}{%
  \enquote{\bibinfo {title} {A criterion for mixed dynamics in two-dimensional
  reversible maps},}\ }%
  \bibfield{journal}{%
  \bibinfo {journal} {Chaos: An Interdisciplinary Journal of Nonlinear
  Science}\ }%
  \textbf{\bibinfo {volume} {31}},\ \bibinfo {pages} {043133} (\bibinfo {year}
  {2021})\BibitemShut{NoStop}%
\bibitem{emelianova2019intersection}%
  \BibitemOpen
  \bibfield{author}{%
  \bibinfo {author} {\bibfnamefont{A.~A.}\ \bibnamefont{Emelianova}}\ and\
  \bibinfo {author} {\bibfnamefont{V.~I.}\ \bibnamefont{Nekorkin}},\ }%
  \bibfield{title}{%
  \enquote{\bibinfo {title} {On the intersection of a chaotic attractor and a
  chaotic repeller in the system of two adaptively coupled phase
  oscillators},}\ }%
  \bibfield{journal}{%
  \bibinfo {journal} {Chaos: An Interdisciplinary Journal of Nonlinear
  Science}\ }%
  \textbf{\bibinfo {volume} {29}},\ \bibinfo {pages} {111102} (\bibinfo {year}
  {2019})\BibitemShut{NoStop}%
\bibitem{ariel2020conservative}%
  \BibitemOpen
  \bibfield{author}{%
  \bibinfo {author} {\bibfnamefont{G.}~\bibnamefont{Ariel}}\ and\ \bibinfo
  {author} {\bibfnamefont{J.}~\bibnamefont{Schiff}},\ }%
  \bibfield{title}{%
  \enquote{\bibinfo {title} {Conservative, dissipative and super-diffusive
  behavior of a particle propelled in a regular flow},}\ }%
  \bibfield{journal}{%
  \bibinfo {journal} {Physica D: Nonlinear Phenomena}\ }%
  \textbf{\bibinfo {volume} {411}},\ \bibinfo {pages} {132584} (\bibinfo {year}
  {2020})\BibitemShut{NoStop}%
\bibitem{gonchenko2020three}%
  \BibitemOpen
  \bibfield{author}{%
  \bibinfo {author} {\bibfnamefont{S.~V.}\ \bibnamefont{Gonchenko}}, \bibinfo
  {author} {\bibfnamefont{A.~S.}\ \bibnamefont{Gonchenko}},\ and\ \bibinfo
  {author} {\bibfnamefont{A.~O.}\ \bibnamefont{Kazakov}},\ }%
  \bibfield{title}{%
  \enquote{\bibinfo {title} {Three types of attractors and mixed dynamics of
  nonholonomic models of rigid body motion},}\ }%
  \bibfield{journal}{%
  \bibinfo {journal} {Proceedings of the Steklov Institute of Mathematics}\ }%
  \textbf{\bibinfo {volume} {308}},\ \bibinfo {pages} {125--140} (\bibinfo
  {year} {2020})\BibitemShut{NoStop}%
\bibitem{kuznetsov2020chaplygin}%
  \BibitemOpen
  \bibfield{author}{%
  \bibinfo {author} {\bibfnamefont{S.~P.}\ \bibnamefont{Kuznetsov}}, \bibinfo
  {author} {\bibfnamefont{V.~P.}\ \bibnamefont{Kruglov}},\ and\ \bibinfo
  {author} {\bibfnamefont{A.~V.}\ \bibnamefont{Borisov}},\ }%
  \bibfield{title}{%
  \enquote{\bibinfo {title} {Chaplygin sleigh in the quadratic potential
  field},}\ }%
  \bibfield{journal}{%
  \bibinfo {journal} {Europhysics Letters}\ }%
  \textbf{\bibinfo {volume} {132}},\ \bibinfo {pages} {20008} (\bibinfo {year}
  {2020})\BibitemShut{NoStop}%
\bibitem{emelianova2020third}%
  \BibitemOpen
  \bibfield{author}{%
  \bibinfo {author} {\bibfnamefont{A.~A.}\ \bibnamefont{Emelianova}}\ and\
  \bibinfo {author} {\bibfnamefont{V.~I.}\ \bibnamefont{Nekorkin}},\ }%
  \bibfield{title}{%
  \enquote{\bibinfo {title} {The third type of chaos in a system of two
  adaptively coupled phase oscillators},}\ }%
  \bibfield{journal}{%
  \bibinfo {journal} {Chaos: An Interdisciplinary Journal of Nonlinear
  Science}\ }%
  \textbf{\bibinfo {volume} {30}},\ \bibinfo {pages} {051105} (\bibinfo {year}
  {2020})\BibitemShut{NoStop}%
\bibitem{emelianova2021emergence}%
  \BibitemOpen
  \bibfield{author}{%
  \bibinfo {author} {\bibfnamefont{A.~A.}\ \bibnamefont{Emelianova}}\ and\
  \bibinfo {author} {\bibfnamefont{V.~I.}\ \bibnamefont{Nekorkin}},\ }%
  \bibfield{title}{%
  \enquote{\bibinfo {title} {Emergence and synchronization of a reversible core
  in a system of forced adaptively coupled {K}uramoto oscillators},}\ }%
  \bibfield{journal}{%
  \bibinfo {journal} {Chaos: An Interdisciplinary Journal of Nonlinear
  Science}\ }%
  \textbf{\bibinfo {volume} {31}},\ \bibinfo {pages} {033102} (\bibinfo {year}
  {2021})\BibitemShut{NoStop}%
\bibitem{Marvel-Mirollo-Strogatz-09}%
  \BibitemOpen
  \bibfield{author}{%
  \bibinfo {author} {\bibfnamefont{S.~A.}\ \bibnamefont{Marvel}}, \bibinfo
  {author} {\bibfnamefont{R.~E.}\ \bibnamefont{Mirollo}},\ and\ \bibinfo
  {author} {\bibfnamefont{S.~H.}\ \bibnamefont{Strogatz}},\ }%
  \bibfield{title}{%
  \enquote{\bibinfo {title} {Phase oscillators with global sinusoidal coupling
  evolve by {M\"o}bius group action},}\ }%
  \bibfield{journal}{%
  \bibinfo {journal} {Chaos}\ }%
  \textbf{\bibinfo {volume} {19}},\ \bibinfo {pages} {043104.} (\bibinfo {year}
  {2009})\BibitemShut{NoStop}%
\bibitem{Gong-Toenjes-Pikovsky-20}%
  \BibitemOpen
  \bibfield{author}{%
  \bibinfo {author} {\bibfnamefont{Ch.~Ch.}\ \bibnamefont{Gong}}, \bibinfo
  {author} {\bibfnamefont{R.}~\bibnamefont{Toenjes}},\ and\ \bibinfo {author}
  {\bibfnamefont{A.}~\bibnamefont{Pikovsky}},\ }%
  \bibfield{title}{%
  \enquote{\bibinfo {title} {Coupled {M}\"obius maps as a tool to model
  {K}uramoto phase synchronization},}\ }%
  \bibfield{journal}{%
  \bibinfo {journal} {Phys. Rev. E}\ }%
  \textbf{\bibinfo {volume} {102}},\ \bibinfo {pages} {022206} (\bibinfo {year}
  {2020})\BibitemShut{NoStop}%
\bibitem{Chigarev-Kazakov-Pikovsky-20}%
  \BibitemOpen
  \bibfield{author}{%
  \bibinfo {author} {\bibfnamefont{V.}~\bibnamefont{Chigarev}}, \bibinfo
  {author} {\bibfnamefont{A.}~\bibnamefont{Kazakov}},\ and\ \bibinfo {author}
  {\bibfnamefont{A.}~\bibnamefont{Pikovsky}},\ }%
  \bibfield{title}{%
  \enquote{\bibinfo {title} {{K}antorovich--{R}ubinstein--{W}asserstein
  distance between overlapping attractor and repeller},}\ }%
  \bibfield{journal}{%
  \bibinfo {journal} {Chaos}\ }%
  \textbf{\bibinfo {volume} {30}},\ \bibinfo {pages} {073114} (\bibinfo {year}
  {2020})\BibitemShut{NoStop}%
\bibitem{Chigarev-Kazakov-Pikovsky-21}%
  \BibitemOpen
  \bibfield{author}{%
  \bibinfo {author} {\bibfnamefont{V.}~\bibnamefont{Chigarev}}, \bibinfo
  {author} {\bibfnamefont{A.}~\bibnamefont{Kazakov}},\ and\ \bibinfo {author}
  {\bibfnamefont{A.}~\bibnamefont{Pikovsky}},\ }%
  \bibfield{title}{%
  \enquote{\bibinfo {title} {Mutual singularities of overlapping attractor and
  repeller},}\ }%
  \bibfield{journal}{%
  \bibinfo {journal} {Chaos}\ }%
  \textbf{\bibinfo {volume} {31}},\ \bibinfo {pages} {083127} (\bibinfo {year}
  {2021})\BibitemShut{NoStop}%
\bibitem{Lai-Tel-11}%
  \BibitemOpen
  \bibfield{author}{%
  \bibinfo {author} {\bibfnamefont{Y.-C.}\ \bibnamefont{Lai}}\ and\ \bibinfo
  {author} {\bibfnamefont{T.}~\bibnamefont{T{\'e}l}},\ }%
  \emph{\bibinfo {title} {Transient Chaos}}\ (\bibinfo {publisher} {Springer},\
  \bibinfo {address} {New York},\ \bibinfo {year} {2011})\BibitemShut{NoStop}%
\bibitem{afraimovich1982bifurcation}%
  \BibitemOpen
  \bibfield{author}{%
  \bibinfo {author} {\bibfnamefont{V.~S.}\ \bibnamefont{Afraimovich}}\ and\
  \bibinfo {author} {\bibfnamefont{L.~P.}\ \bibnamefont{Shilnikov}},\ }%
  \bibfield{title}{%
  \enquote{\bibinfo {title} {On a bifurcation of codimension 1 leading to the
  appearance of a countable set of tori},}\ }%
  \bibfield{journal}{%
  \bibinfo {journal} {Doklady Akademii Nauk}\ }%
  \textbf{\bibinfo {volume} {262}},\ \bibinfo {pages} {777--780} (\bibinfo
  {year} {1982})\BibitemShut{NoStop}%
\bibitem{Grebogi-Ott-Yorke-83c}%
  \BibitemOpen
  \bibfield{author}{%
  \bibinfo {author} {\bibfnamefont{C.}~\bibnamefont{Grebogi}}, \bibinfo
  {author} {\bibfnamefont{E.}~\bibnamefont{Ott}},\ and\ \bibinfo {author}
  {\bibfnamefont{J.~A.}\ \bibnamefont{Yorke}},\ }%
  \bibfield{title}{%
  \enquote{\bibinfo {title} {{Fractal basin boundaries, long-lived chaotic
  transients, and unstable-unstable pair bifurcation}},}\ }%
  \bibfield{journal}{%
  \bibinfo {journal} {Phys. Rev. Lett.}\ }%
  \textbf{\bibinfo {volume} {50}},\ \bibinfo {pages} {935--938} (\bibinfo
  {year} {1983})\BibitemShut{NoStop}%
\bibitem{Grebogi-Ott-Yorke-85}%
  \BibitemOpen
  \bibfield{author}{%
  \bibinfo {author} {\bibfnamefont{C.}~\bibnamefont{Grebogi}}, \bibinfo
  {author} {\bibfnamefont{E.}~\bibnamefont{Ott}},\ and\ \bibinfo {author}
  {\bibfnamefont{J.~A.}\ \bibnamefont{Yorke}},\ }%
  \bibfield{title}{%
  \enquote{\bibinfo {title} {{Super-persistent chaotic transients}},}\ }%
  \bibfield{journal}{%
  \bibinfo {journal} {Ergod. Theory Dynam. Sys.}\ }%
  \textbf{\bibinfo {volume} {5}},\ \bibinfo {pages} {341--372} (\bibinfo {year}
  {1985})\BibitemShut{NoStop}%
\bibitem{olicon2020critical}%
  \BibitemOpen
  \bibfield{author}{%
  \bibinfo {author} {\bibfnamefont{G.}~\bibnamefont{Olicon~Mendez}},\ }%
  \emph{\bibinfo {title} {Critical behaviour of random diffeomorphisms:
  quasi-stationary measures and escape times}},\ Ph.D. thesis,\ \bibinfo
  {school} {Imperial College London} (\bibinfo {year}
  {2020})\BibitemShut{NoStop}%
\bibitem{KOSTELICH199781}%
  \BibitemOpen
  \bibfield{author}{%
  \bibinfo {author} {\bibfnamefont{E.~J.}\ \bibnamefont{Kostelich}}, \bibinfo
  {author} {\bibfnamefont{I.}~\bibnamefont{Kan}}, \bibinfo {author}
  {\bibfnamefont{C.}~\bibnamefont{Grebogi}}, \bibinfo {author}
  {\bibfnamefont{E.}~\bibnamefont{Ott}},\ and\ \bibinfo {author}
  {\bibfnamefont{J.~A.}\ \bibnamefont{Yorke}},\ }%
  \bibfield{title}{%
  \enquote{\bibinfo {title} {Unstable dimension variability: A source of
  nonhyperbolicity in chaotic systems},}\ }%
  \bibfield{journal}{%
  \bibinfo {journal} {Physica D}\ }%
  \textbf{\bibinfo {volume} {109}},\ \bibinfo {pages} {81--90} (\bibinfo {year}
  {1997})\BibitemShut{NoStop}%
\bibitem{Das-Yorke-17}%
  \BibitemOpen
  \bibfield{author}{%
  \bibinfo {author} {\bibfnamefont{S.}~\bibnamefont{Das}}\ and\ \bibinfo
  {author} {\bibfnamefont{J.~A.}\ \bibnamefont{Yorke}},\ }%
  \bibfield{title}{%
  \enquote{\bibinfo {title} {Multichaos from quasiperiodicity},}\ }%
  \bibfield{journal}{%
  \bibinfo {journal} {SIAM Journal on Applied Dynamical Systems}\ }%
  \textbf{\bibinfo {volume} {16}},\ \bibinfo {pages} {2196--2212} (\bibinfo
  {year} {2017})\BibitemShut{NoStop}%
\bibitem{alligood2006crossing}%
  \BibitemOpen
  \bibfield{author}{%
  \bibinfo {author} {\bibfnamefont{K.~T.}\ \bibnamefont{Alligood}}, \bibinfo
  {author} {\bibfnamefont{E.}~\bibnamefont{Sander}},\ and\ \bibinfo {author}
  {\bibfnamefont{J.~A.}\ \bibnamefont{Yorke}},\ }%
  \bibfield{title}{%
  \enquote{\bibinfo {title} {Crossing bifurcations and unstable dimension
  variability},}\ }%
  \bibfield{journal}{%
  \bibinfo {journal} {Physical Review Letters}\ }%
  \textbf{\bibinfo {volume} {96}},\ \bibinfo {pages} {244103} (\bibinfo {year}
  {2006})\BibitemShut{NoStop}%
\bibitem{Li_2017}%
  \BibitemOpen
  \bibfield{author}{%
  \bibinfo {author} {\bibfnamefont{D.}~\bibnamefont{Li}},\ }%
  \bibfield{title}{%
  \enquote{\bibinfo {title} {Homoclinic bifurcations that give rise to
  heterodimensional cycles near a saddle-focus equilibrium1},}\ }%
  \bibfield{journal}{%
  \bibinfo {journal} {Nonlinearity}\ }%
  \textbf{\bibinfo {volume} {30}},\ \bibinfo {pages} {173} (\bibinfo {year}
  {2016})\BibitemShut{NoStop}%
\bibitem{Li_17}%
  \BibitemOpen
  \bibfield{author}{%
  \bibinfo {author} {\bibfnamefont{D.}~\bibnamefont{Li}}\ and\ \bibinfo
  {author} {\bibfnamefont{D.}~\bibnamefont{Turaev}},\ }%
  \bibfield{title}{%
  \enquote{\bibinfo {title} {Existence of heterodimensional cycles near
  shilnikov loops in systems with a $\mathbb{Z}_2$ symmetry},}\ }%
  \bibfield{journal}{%
  \bibinfo {journal} {Discrete and Continuous Dynamical Systems}\ }%
  \textbf{\bibinfo {volume} {37}},\ \bibinfo {pages} {4399--4437} (\bibinfo
  {year} {2017})\BibitemShut{NoStop}%
\bibitem{hittmeyer2018existence}%
  \BibitemOpen
  \bibfield{author}{%
  \bibinfo {author} {\bibfnamefont{S.}~\bibnamefont{Hittmeyer}}, \bibinfo
  {author} {\bibfnamefont{B.}~\bibnamefont{Krauskopf}}, \bibinfo {author}
  {\bibfnamefont{H.~M.}\ \bibnamefont{Osinga}},\ and\ \bibinfo {author}
  {\bibfnamefont{K.}~\bibnamefont{Shinohara}},\ }%
  \bibfield{title}{%
  \enquote{\bibinfo {title} {Existence of blenders in a {H}{\'e}non-like
  family: Geometric insights from invariant manifold computations},}\ }%
  \bibfield{journal}{%
  \bibinfo {journal} {Nonlinearity}\ }%
  \textbf{\bibinfo {volume} {31}},\ \bibinfo {pages} {R239} (\bibinfo {year}
  {2018})\BibitemShut{NoStop}%
\bibitem{Li_2020}%
  \BibitemOpen
  \bibfield{author}{%
  \bibinfo {author} {\bibfnamefont{D.}~\bibnamefont{Li}}\ and\ \bibinfo
  {author} {\bibfnamefont{D.}~\bibnamefont{Turaev}},\ }%
  \bibfield{title}{%
  \enquote{\bibinfo {title} {Persistent heterodimensional cycles in periodic
  perturbations of {L}orenz-like attractors},}\ }%
  \bibfield{journal}{%
  \bibinfo {journal} {Nonlinearity}\ }%
  \textbf{\bibinfo {volume} {33}},\ \bibinfo {pages} {971} (\bibinfo {year}
  {2020})\BibitemShut{NoStop}%
\bibitem{hittmeyer2020identify}%
  \BibitemOpen
  \bibfield{author}{%
  \bibinfo {author} {\bibfnamefont{S.}~\bibnamefont{Hittmeyer}}, \bibinfo
  {author} {\bibfnamefont{B.}~\bibnamefont{Krauskopf}}, \bibinfo {author}
  {\bibfnamefont{H.~M.}\ \bibnamefont{Osinga}},\ and\ \bibinfo {author}
  {\bibfnamefont{K.}~\bibnamefont{Shinohara}},\ }%
  \bibfield{title}{%
  \enquote{\bibinfo {title} {How to identify a hyperbolic set as a blender},}\
  }%
  \bibfield{journal}{%
  \bibinfo {journal} {Discrete \& Continuous Dynamical Systems}\ }%
  \textbf{\bibinfo {volume} {40}},\ \bibinfo {pages} {6815} (\bibinfo {year}
  {2020})\BibitemShut{NoStop}%
\bibitem{Hammerlindl_22}%
  \BibitemOpen
  \bibfield{author}{%
  \bibinfo {author} {\bibfnamefont{A.}~\bibnamefont{Hammerlindl}}, \bibinfo
  {author} {\bibfnamefont{B.}~\bibnamefont{Krauskopf}}, \bibinfo {author}
  {\bibfnamefont{G}~\bibnamefont{Mason}},\ and\ \bibinfo {author}
  {\bibfnamefont{Osinga~H.}\ \bibnamefont{M.}},\ }%
  \bibfield{title}{%
  \enquote{\bibinfo {title} {Determining the global manifold structure of a
  continuous-time heterodimensional cycle},}\ }%
  \bibfield{journal}{%
  \bibinfo {journal} {Journal of Computational Dynamics}\ }%
  \textbf{\bibinfo {volume} {9}},\ \bibinfo {pages} {393--419} (\bibinfo {year}
  {2022})\BibitemShut{NoStop}%
\bibitem{Kuptsov_2013}%
  \BibitemOpen
  \bibfield{author}{%
  \bibinfo {author} {\bibfnamefont{P.~V.}\ \bibnamefont{Kuptsov}},\ }%
  \bibfield{title}{%
  \enquote{\bibinfo {title} {Violation of hyperbolicity via unstable dimension
  variability in a chain with local hyperbolic chaotic attractors},}\ }%
  \bibfield{journal}{%
  \bibinfo {journal} {J. Phys. A: Math. Theor.}\ }%
  \textbf{\bibinfo {volume} {46}},\ \bibinfo {pages} {254016} (\bibinfo {year}
  {2013})\BibitemShut{NoStop}%
\bibitem{Vallejo-Sanjuan-16}%
  \BibitemOpen
  \bibfield{author}{%
  \bibinfo {author} {\bibfnamefont{J.~C.}\ \bibnamefont{{Vallejo}}}\ and\
  \bibinfo {author} {\bibfnamefont{M.~A.~F.}\ \bibnamefont{{Sanjuan}}},\ }%
  \bibfield{title}{%
  \enquote{\bibinfo {title} {Role of dark matter haloes on the predictability
  of computed orbits},}\ }%
  \bibfield{journal}{%
  \bibinfo {journal} {Astronomy \& Astrophysics}\ }%
  \textbf{\bibinfo {volume} {595}},\ \bibinfo {eid} {A68} (\bibinfo {year}
  {2016})\BibitemShut{NoStop}%
\bibitem{Esashi_etal-18}%
  \BibitemOpen
  \bibfield{author}{%
  \bibinfo {author} {\bibfnamefont{K.}~\bibnamefont{Esashi}}, \bibinfo {author}
  {\bibfnamefont{T.}~\bibnamefont{Onozaki}}, \bibinfo {author}
  {\bibfnamefont{Y.}~\bibnamefont{Saiki}},\ and\ \bibinfo {author}
  {\bibfnamefont{Y.}~\bibnamefont{Sato}},\ }%
  \bibfield{title}{%
  \enquote{\bibinfo {title} {Intermittent transition between synchronization
  and desynchronization in multi-regional business cycles},}\ }%
  \bibfield{journal}{%
  \bibinfo {journal} {Structural Change and Economic Dynamics}\ }%
  \textbf{\bibinfo {volume} {44}},\ \bibinfo {pages} {68--76} (\bibinfo {year}
  {2018})\BibitemShut{NoStop}%
\bibitem{Pikovsky-Politi-16}%
  \BibitemOpen
  \bibfield{author}{%
  \bibinfo {author} {\bibfnamefont{A.}~\bibnamefont{Pikovsky}}\ and\ \bibinfo
  {author} {\bibfnamefont{A.}~\bibnamefont{Politi}},\ }%
  \emph{\bibinfo {title} {Lyapunov Exponents. A Tool to Explore Complex
  Dynamics}}\ (\bibinfo {publisher} {Cambridge University Press},\ \bibinfo
  {address} {Cambridge},\ \bibinfo {year} {2016})\BibitemShut{NoStop}%
\end{thebibliography}

%

\appendix
\section{Properties of a M\"obius map}
\label{sec:app}
It is convenient to define a M\"obius map as a map of a circle $\psi\to\bar{\psi}=M(\e,\Phi,\Psi)\psi$, where $0\leq \psi<2\pi$:
\begin{equation}
e^{i(\bar{\psi}-\Phi)}=\frac{\e+e^{i(\psi-\Psi)}}{\e e^{i(\psi-\Psi)}+1}\;.
\label{eq:mmf}
\end{equation}
Using identity $\exp[i\alpha]=(1+i\tan\frac{\alpha}{2})(1-i\tan\frac{\alpha}{2})^{-1}$, we can rewrite the map in a real form as
\begin{equation}
\tan\left(\frac{\bar\psi-\Phi}{2}\right)=\frac{1-\e}{1+\e}
\tan\left(\frac{\psi-\Psi}{2}\right)\;.
\end{equation}
This form is, however, not convenient for numerical computation of the rotation number. If we first rewrite \eqref{eq:mmf} as
\[
e^{i(\bar\psi-\Phi)}=e^{i(\psi-\Psi)}
\frac{1+\e e^{-i(\psi-\Psi)}}{1+\e e^{i(\psi-\Psi)}}\;,
\]
then another representation
\begin{equation}
\bar\psi=\psi+\Phi-\Psi-2\arctan\frac{\e\sin(\psi-\Psi)}{1+\e\cos(\psi-\Psi)}
\label{eq:mmf2}
\end{equation}
follows, which corresponds to Eq.~\eqref{eq:bmm}.

For the next calculations, yet another reformulation
\begin{equation}
\bar\psi=M(\alpha,z)\psi=e^{i\alpha}\frac{z+e^{i\psi}}{1+z^*e^{i\psi}}
\label{eq:mmf3}
\end{equation}
is suitable, where $z=\e \exp[i\Psi]$ and $\alpha=\Phi-\Psi$. A direct application of \eqref{eq:mmf3} shows that a composition
of two M\"obius maps is again a M\"obius map
\begin{equation}
\begin{gathered}
M(\alpha_2,z_2)M(\alpha_1,z_1)=M(\alpha_1+\alpha_2+Q, Z)\;,\\
 Z=\frac{z_1+\left(z_2e^{-i\alpha_1}\right)}{1+z_1^*\left(z_2e^{-i\alpha_1}\right)}\;,\qquad
e^{iQ}=\frac{1+z_1^*\left(z_2e^{-i\alpha_1}\right)}{1+z_1\left(z_2e^{-i\alpha_1}\right)^*}\;.
\end{gathered}
\label{eq:2mm}
\end{equation}

Finally, we show that a M\"obius map can be transformed (in a certain range of parameters) to a circle shift.
The transformation itself is a M\"obius transformation with certain complex parameter $s=\kappa\exp[i\xi]$:
\begin{equation}
e^{i\psi}=\frac{s+e^{i\phi}}{1+s^*e^{i\phi}}\;.
\end{equation}
Substituting this in \eqref{eq:mmf3} one can show that the transformation for $\phi$ has the form $\bar\phi=\phi+\Theta$, provided
\begin{equation}
\kappa=\e^{-1}\left(\sqrt{\sin^2\frac{\alpha}{2}-\e^2}-\sin\frac{\alpha}{2}\right),\quad \xi=\frac{\pi}{2}-\frac{\alpha}{2}-\Psi\;.
\end{equation}
One can see that such a transformation is possible if $\e<|\sin\frac{\alpha}{2}|$, otherwise the M\"obius transformation has a stable and an unstable fixed points.
The rotation number is
\begin{equation}
\Theta=2\arctan\left(\tan\frac{\alpha}{2}\sqrt{1-\frac{\e^2}{\sin^2\frac{\alpha}{2}}}\right)\;.
\label{eq:at}
\end{equation}
This conjugation to a circle shift means that the M\"obius map has only one Arnold tongue $\e\geq |\sin\frac{\alpha}{2}|$, outside of which the rotation number is a smooth function of parameters as follows from relation \eqref{eq:at}.
\end{document}